\documentclass{aa}
\usepackage{graphicx}
\usepackage{txfonts}
\usepackage{longtable,lscape}
\usepackage{natbib}
\bibpunct{(}{)}{;}{a}{}{,}

\def\be{\begin{equation}}
\def\ee{\end{equation}}
\def\kms{{\rm\,km\,s^{-1}}}
\def\kmskpc{{\rm\,km\,s^{-1}\,{kpc}^{-1}}}
\def\pc{{\rm\,pc}}

\def\Myr{{\rm\,Myr}}
\def\Gyr{{\rm\,Gyr}}
\def\deg{{^\circ}}

\def\dex{{\rm\,dex}}
\begin{document}
   \title{Origin and evolution of moving groups}
 \subtitle{I. Characterization in the observational kinematic-age-metallicity space}
   \author{T. Antoja
      \and F. Figueras
      \and D. Fern\'andez
      \and J. Torra
          }
   \offprints{T. Antoja,\\
   \email{tantoja@am.ub.es}}
   \institute{Departament d'Astronomia i Meteorologia and IEEC-UB,
     Institut de Ci\`encies del Cosmos de la Universitat de Barcelona,
     Mart\'i i Franqu\`es, 1, E-08028 Barcelona, Spain}
   \date{Received 06 February 2008 / Accepted 05 August 2008}
   \abstract
{Recent studies have suggested that moving groups have a dynamic or ``resonant'' origin. Under this hypothesis, these kinematic structures become a powerful tool for studying the large-scale structure and dynamics of the Milky Way.}
{Here we aim to characterize these structures in the $U$--$V$--$age$--$[Fe/H]$ space and establish observational constraints that will allow us to study their origin and evolution.}
{We apply multiscale techniques -- wavelet denoising (WD)-- to an extensive compendium of more than 24000 stars in the solar neighbourhood with the best available astrometric, photometric and spectroscopic data.}
{We confirm  that the dominant structures in the $U$--$V$ plane are the branches of Sirius, Coma Berenices, Hyades-Pleiades and Hercules.  These branches are nearly equidistant in this kinematic plane and they show a negative slope. The abrupt drops in the velocity density distribution are characterized. We find a certain dependence of these kinematic structures on Galactic position with a significant change of contrast among substructures inside the branches. A large spread of ages is observed for all branches. The Hercules branch is detected in all subsamples with ages older than $\sim 2\Gyr$ and the set of the other three branches is well established for stars $>400 \Myr$. The age-metallicity relation of each branch is examined and the relation between kinematics and metallicity is studied.}
{Not all of these observational constraints are successfully explained by the recent models proposed for the formation of such kinematic structures. Simulations incorporating stellar ages and metallicities are essential for future studies. The comparison of the observed and simulated distributions obtained by WD will provide a physical interpretation of the existence of the branches in terms of local or large-scale dynamics.}
\maketitle
\section{Introduction}\label{introduction}

After the pioneering work of \citet{proctor1869}, \citet{kapteyn05} and \citet{lindblad25}, Eggen established the spatial and kinematic properties of several {\it stellar streams} - hereafter classic moving groups - formed by stars with similar kinematics in the solar neighbourhood (see \citealt{eggen96} and references therein). Since these structures share their kinematics with certain open clusters, he worked exhaustively on the hypothesis that moving groups are a result of the dispersion (resulting from both external and internal causes)
of stellar clusters. 

 The advent of Hipparcos astrometric data led to the definitive establishment of the existence of moving groups and to the recognition of substructures within them \citep{chereul98,asiain99a}. In addition, together with \citet{dehnen98}, these authors first attempted to evaluate the evolutionary state of the members of moving groups, which is necessary in order to establish their origin and evolution. \citet{dehnen98} used subsamples composed of stars with different spectral types to study the velocity distribution of old and young stars. As no photometric data were available, \citet{chereul98} could only use what they called ``palliative'' ages to show that the age distribution of the stars in moving groups seems to be similar to that observed for their whole sample. More precise ages, derived using Str{\"o}mgren photometry, were used by \citet{asiain99b}, who observed that the velocity dispersion of several substructures within the Pleiades moving group is compatible with that expected for the evolution of a stellar complex \citep{efremov88}. This conclusion was restricted to groups of stars with ages of up to about $1\Gyr$, as both differential Galactic rotation and disk heating would have dispersed older groups among the field stars. Using an adaptive kernel and wavelet transform (WT) analysis, \citet{skuljan99} studied a sample of 4000 Hipparcos stars and found that the distribution function in the $U$--$V$ plane is characterized by a few branches that are diagonal, parallel and roughly equidistant. Later on, \citet{famaey05} used a maximum-likelihood method based on a Bayesian approach to divide a sample of giant stars into several kinematic groups. A very wide range of ages for each of these structures in their corresponding H-R diagram was observed.

The first theoretical arguments in favour of a different dynamic origin of moving groups were put forward by \citet{mayor72} and \citet{kalnajs91}. Following this latter author, \citet{dehnen98} pointed out that orbital resonances could be the cause of the existence of most moving groups observed in the solar neighbourhood. \citet{skuljan99} related the origin of the branches to the Galactic spiral structure, or to some other global characteristics of the Galactic potential, combined with the initial velocities of the stars. Nowadays, these dynamic or ``resonant'' mechanisms are proposed to be the most plausible explanation for the main moving groups. Other minor kinematic structures in the solar neighbourhood, such as HR1614, do seem to be remnants of a dispersed star-forming event \citep{desilva07} and it has been proposed that other structures are related to accretion events in the Galaxy \citep{helmi06}.

The work by \citet{dehnen00} and \citet{fux00,fux01}, among others, focused on the origin of the Hercules structure and its connection to Galactic bar resonances. Furthermore, by numerically integrating test particle orbits, \citet{desimone04} showed that stochastic spiral density waves can produce kinematic structures similar to those found by \citet{skuljan99}. \citet{quillen05} and \citet{chakrabarty07} have recently gone one step further and looked for the relation between moving groups and resonances of the non-axisymmetric component of the Galactic potential, i.e. the Galactic bar and spiral structure. They have shown that by using several combinations of the parameters adopted to characterize this potential they are able to reproduce structures in the kinematic plane similar to those actually observed. To restrict the free parameters of the Galactic potential, all this recent work requires new observational constraints such as a characterization of the observed kinematic structures in the $U$--$V$ plane in terms of their evolutionary state, their chemical composition or their possible dependence on Galactic position. 

Nowadays, two important observational contributions provide new material to complement the Hipparcos and Tycho astrometric data: i) the CORAVEL radial velocity data for a significant number of late-type stars belonging to the Hipparcos catalogue (\citealt{nordstrom04} for dwarf stars and \citealt{famaey05} for giant stars) and ii) the uvby--$\beta$ survey of FGK dwarf stars, which have allowed the derivation of ages and metallicities \citep{nordstrom04}. Lastly, data on OBA-type stars \citep{asiain99a,torra00} and M dwarfs \citep{reid02,bochanski05} complete an extensive sample of more than 24000 stars ready to be used for the characterization of moving groups. It is necessary to complement these new data with more powerful statistical tools to interpret both the observed and simulated data in the 4-dimensional space of $U$--$V$--$age$--$[Fe/H]$.

In this paper, we characterize the observed kinematic structures in $U$--$V$--$age$-$[Fe/H]$ space by applying new statistical techniques with the aim of establishing observational constraints that will help to reveal their origin and evolution. Section \ref{data} of this paper presents the observational data we use together with their precisions and possible biases. Section \ref{method} contains a description of the statistical method of the wavelet denoising (WD) we apply. Section \ref{velocity} characterizes in depth the structures in the velocity plane. Then, Sect. \ref{age} and Sect. \ref{feh} analyse the age and metallicity distributions of these structures. Finally, the main outcomes of the work and perspectives for the future are summarized in Sect. \ref{conclusions}. Subsequent papers will further this research by using these constraints in combination with test particle simulations with a suitable Galactic potential model.

\section{Compilation of available data}\label{data}

\begin{table*}
  \caption{Number of stars with kinematic data, age and metallicity from each catalogue and for the total sample and error information.} 
   \label{tab.data}
\centering
\begin{tabular}{llllll}\hline\hline
Data                       & OBA (1, 2) & FGK (3)&M (4, 5)           &KM giants (6)&total      \\\hline
$U$ $V$ $W$                & 4283       & 13257  & 863               & 5787        & \bf 24190 \\
$U$ $V$ $W$ $age$          &3977        &11215   &863\footnotemark[2]&0 &\bf 16055  \\
$U$ $V$ $W$ $[Fe/H]$       &0 &13109   &0        &0 &\bf 13109  \\
$U$ $V$ $W$ $age$ $[Fe/H]$ &0 &11215   &0        &0&\bf 11215  \\\hline
$\epsilon_{U, V, W}\leq 2 \kms$             &$21\%$       &$54 \%$      &...&$17\%$       &...\\
$\bar \epsilon_U$, $\bar \epsilon_V$, $\bar \epsilon_W$  ($\kms$)&4.5, 4.4, 3.9&2.0, 1.9, 1.6&...&4.0, 3.5, 2.8&...\\\hline
$\epsilon_{age}\leq 30\%$                   &$32\%$       &$50 \%$      &...&...        &...\\\hline
\multicolumn{6}{l}{\scriptsize References. (1) \citet{asiain99a}; (2) \citet{torra00}; (3)\citet{nordstrom04}; (4) \citet{reid02};}\\
\multicolumn{6}{l}{\scriptsize (5) \citet{bochanski05}; (6) \citet{famaey05}.}\\
\end{tabular}
\end{table*}

The observational data is compiled from several recent catalogues and is restricted to samples with precise radial velocities and astrometric data, as these are necessary to compute the $U$, $V$ and $W$ heliocentric velocity components\footnote{$U$ is the velocity component (positive) towards the Galactic anti-centre, $V$ is (positive) in the direction of the Galactic rotation and $W$ (positive) toward the north Galactic pole.}. In order to study the evolutionary state of the kinematic structures, stellar ages and metallicities are required. For this reason, we aim to cover a wide range of ages as well as spectral types and luminosity classes. Therefore, catalogues where ages are estimated using photometric calibrations as well as catalogues containing parameters closely related to age (H$_\alpha$ equivalent width\footnote{Equivalent width of the H$_\alpha$ emission line can be used in this case as an indicator of the chromospheric activity and, consequently, as age information.}) are considered. Table \ref{tab.data} shows the data available in each catalogue and information on parameter precision. Altogether our sample contains 24190 stars and the age of over 16000 stars can be used. The following is a detailed description of the sample:

\begin{itemize}
\item {\it OBA-type Stars}: Sample of 4283 stars compiled by \citet{asiain99a} (2025 A-type stars) and \citet{torra00} (2258 O- and B-type stars) with astrometric data from Hipparcos and stellar physical parameters and ages computed from Str{\"o}mgren photometry. Most of the radial velocities came from the radial velocity survey of Hipparcos early-type stars \citep{fehrenbach87,grenier99} planned to complement the kinematically unbiased CORAVEL database for late-type stars. For the OB stars, when both trigonometric and photometric distances were available, the one with smaller relative error was used. Relative error on photometric distance ranges from 14 to 25\%, depending on spectral type and luminosity class. In the compilation of A stars, the Hipparcos distance was adopted when the relative error in the parallax was $\sigma_\pi/\pi<20$\% and the photometric distance otherwise.

\item {\it FGK-type Dwarfs}: Sample of 13257 stars selected from the 16682 stars in the catalogue of the Geneva-Copenhagen survey \citep{nordstrom04} with kinematic data, ages and metallicities. Proper motions come from Tycho-2 and most of their radial velocities were obtained from CORAVEL. As detailed in \cite{nordstrom04}, the Hipparcos parallax was used when $\sigma_\pi/\pi<13$\% and otherwise the photometric distance was employed. Ages and $[Fe/H]$ were calculated from uvby-$\beta$ photometry. The Bayesian estimation method in \citet{jorgensen05} was used for age computation\footnote{In the original catalogue, ages are rounded up to the nearest 0.1 $\Gyr$ but in this study, ages (and upper and lower 1$\sigma$ confidence limits) rounded up to 0.01$\Gyr$ are used \citep{jorgensen06}.}.

\item {\it M-type Dwarfs}: A set of 863 stars selected from the spectroscopic surveys in \citet{bochanski05} (428 stars) and \citet{reid02} (435 stars). The former catalogue gives spectroscopic distances obtained from the TiO$_5$ band, radial velocities from cross-correlation in the 6000-7400 $\AA$ range and USNO-B proper motions \citep{monet03}. \citet{reid02} use mainly proper motions and trigonometric distances from the Hipparcos catalogue and radial velocity data from  \citet{gizis02} and \citet{delfosse99}. Both surveys give the equivalent width of the H$_\alpha$ line, which is indicative of the magnetic activity in the chromosphere and is directly related to stellar age. Following the criteria in \citet{bochanski05}, this sample can be split into two subsamples: active stars (182 young stars) and non-active stars (681 older stars) depending on whether the width of the H$\alpha$ line is $> 1\AA$ or $\le 1 \AA$, respectively. 

\item {\it KM-type Giants}: A group of 5787 giant stars from the catalogue by \citet{famaey05}. They belong to the Hipparcos catalogue and have radial velocities given in the CORAVEL database and proper motions taken from Tycho-2. Their distances were computed using the maximum likelihood method developed by \citet{luri96}, which is based on a Bayesian and parametric approach.

\end{itemize}

\begin{figure}
 \includegraphics[width=0.47\textwidth]{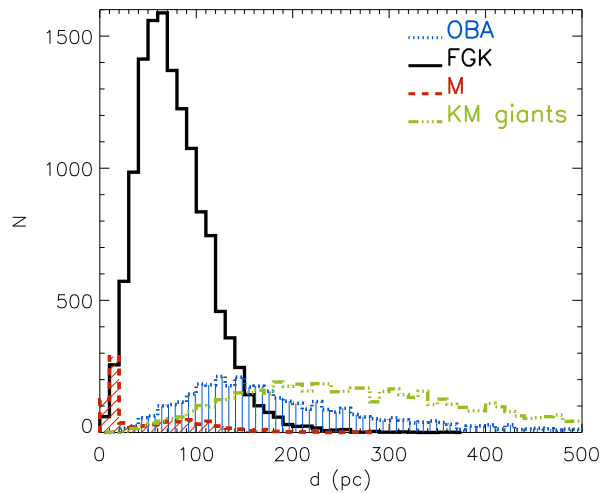}
    \caption{Distance distributions of stars from the catalogues that make up the sample.}
    \label{hisdist}
\end{figure}

Figure \ref{hisdist} shows the distance distribution of the stars from the different catalogues that make up our sample. As expected, the volume of space sampled is highly dependent on the spectral type, which will be taken into account in our subsequent analysis. Distance error distributions and possible biases in the distance adopted are discussed in the publications associated with each catalogue. As mentioned by \citet{skuljan99}, large distance errors could lead to features in the kinematic $U$--$V$ plane being artificially radially elongated (relative to the point $(U,V)=(0,0)$). We have checked that in our sample, distance errors are smaller than 25\%, except for 4\% of stars, all of them KM giants, which in any case exceed 40\%.

No systematic bias in the photometric distances is expected \citep[see e.g.][]{nordstrom04}. For trigonometric distances (1/$\pi$), \citet{brown97} showed that a symmetric error law for parallaxes such as a Gaussian results in a non-symmetric or systematic error in distances, which leads to a biased overestimated distance distribution. This relative bias is shown to be $\approx (\sigma_{\pi}/\pi)^2$ \citep{arenou99}. This expression gives a maximum bias of 2\% for stars with trigonometric distances of the FGK subsample --all of them with $\sigma_{\pi}/\pi\le 0.13$--, which accounts for about $\sim$75\% of all the trigonometric parallaxes in our sample. Moreover, the bias for the whole sample is estimated to be always less than 6\%. The maximum likelihood method used to derive distances for the KM giants corrects this and other systematic trends \citep[see Fig. 8 in][]{famaey05}. However, as discussed in Sect. \ref{spectral}, other kinds of biases caused by the a priori parametrization of the kinematic distribution function cannot be ruled out.

Heliocentric velocities and their errors have been recalculated when the necessary data were available. For the OBA-type stars and KM-type giant stars, most of which have distances $> 100 \pc$, the velocities have been corrected for Galactic differential rotation with values for Oort's constants of: $A=14.82 \kmskpc$ and $B=-12.37 \kmskpc$ \citep{feast97}.

At this point it is of utmost importance to ascertain the possible kinematic biases of our sample due to selection effects. As stated by \citet{binney97}, high proper motion stars were preferred in pre-Hipparcos radial velocity programmes, which leaded to kinematically severely biased samples. During the last two decades, specific observational programmes for radial velocities were undertaken in parallel with the Hipparcos mission (CORAVEL and \citealt{fehrenbach87}) to complete kinematic data for stars of the Hipparcos survey. Most of the stars in the subsamples of FGK dwarfs and KM giants belong to this survey. Consequently, this bias is expected to be suppressed for 79\% of the stars of our whole sample. For the OB stars, \citet{torra00} discussed in detail a small kinematic bias in their subsample (see Fig. 3 therein). Numerical simulations allowed these authors to demonstrate that it had negligible effects on their kinematic analysis. Regarding the A-type stars, some of the photometric programmes on which the \citet{asiain99a} compilation is based favoured stars with known pre-Hipparcos radial velocity data. Although a slight bias is expected for this subsample, it only represents about 8\% of the whole sample. Finally, as was evaluated by \citet{reid02}, their M dwarfs catalogue --a volume limited sample-- shows no evidence of any systematic bias (see Fig. 4 therein). On the other hand, the early M-type stars from \citet{bochanski05} were photometrically selected by color and magnitude and thus no kinematic bias from the selection process is expected. Thus, all these considerations allow us to assume confidently that no sample selection bias affects our kinematic study in the next sections.

\begin{figure}
\includegraphics[width=0.47\textwidth]{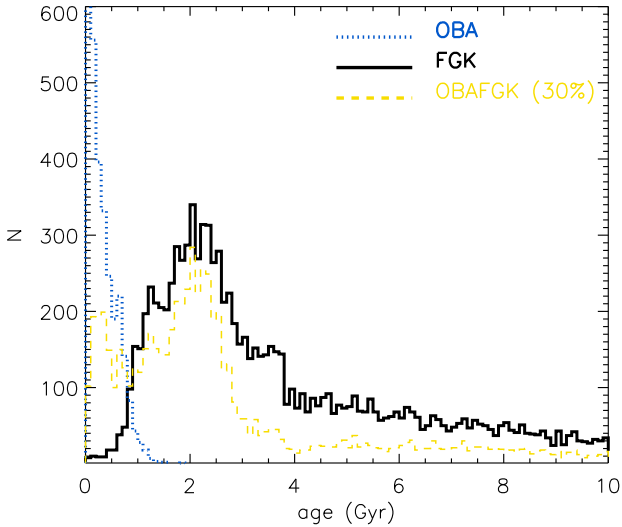}
	\caption{Age distribution for the OBA-type stars (dotted line) and for the FGK-type stars (solid line). The dashed line shows the histogram for stars with relative errors in age of less than 30\%.}
    \label{hisage}
\end{figure}

Figure \ref{hisage} shows the age distribution for the OBA- and FGK-type dwarf stars in our sample. The precision of this parameter deserves special attention. The vast majority of the OBA-type stars are main sequence or moderately evolved stars and, therefore, a reliable estimate for their ages is available from uvby-$\beta$ photometry (see the error distribution in Fig. 1b of \citealt{asiain99a} and Fig. 7 of \citealt{torra00}). As stated by \citet{torra00}, an overestimation in age as large as 30-50\% is expected for highly rotating stars, that are a substantial fraction of the OB main-sequence stars. This, however, implies an absolute bias of always less than 50 $\Myr$ whose effect in our age-kinematic analysis undertaken in Sect. \ref{age} is almost negligible. Figure 15 in \citet{nordstrom04} shows the distribution of upper and lower 1$\sigma$ re\-la\-ti\-ve errors in age for the FGK sample. Large relative errors in age lead to the evolutionary state of late-type stars being largely undetermined. Furthermore, certain biases in the age estimation would have non-negligible consequences for our study: dependence on the evolutionary models used or biases that arise in certain regions of the HR diagram\footnote{For instance, \citet{pont04} demonstrate that the isochrone dating method is subject to significant biases that arise inevitably in regions of the HR diagram where the effects of age on the atmospheric parameters become small. \citet{haywood06} recently studied this in depth and showed how biases, similar to those expected in real samples, greatly influence the physical interpretation of the Galactic age-metallicity distribution.}.

Using improved calibrations, \citet{holmberg07} recomputed ages and error estimates for the \citet{nordstrom04} sample. They state that the differences between old (used here) and new values\footnote{They were not available when the present work was completed.} are insignificant and are much smaller than the estimated individual uncertainties. Furthermore, they find an agreement between the new ages and the independent age determination by \citet{takeda07} (see Fig. 19 of \citealt{holmberg07}). All these results allow us to be slightly more confident of the age values and their uncertainties. However, a cut-off by error for this parameter allows us to work with the more reliable ages. The distribution of stars with relative errors in their ages\footnote{\citet{nordstrom04} provides upper and lower limits for error estimates. Here we consider that a star has $\epsilon_{age}\leq 30\%$ if this is true for both limits.} of less than 30\% is shown in Fig. \ref{hisage}. This condition is fulfilled by 32\% and 50\% of stars in the OBA and FGK samples, respectively. It has to be kept in mind that employing this cut-off results in a change in the content of the working sample (e.g. uncertainties in isochrone ages increase when less massive main sequence stars are considered\footnote{For a discussion of the dependence of the degree of completeness of their sample on the cut-off by error in ages see \citet{nordstrom04}.}). The most significant feature of this age histogram is the peak around $2 \Gyr$. This peak results from a combination of selection effects in the sample in \citet{nordstrom04}: apparent magnitude, spectral type, etc. However, the lack of stars younger than $\sim 1 \Gyr$ --a direct result of the blue cut-off at $b-y=0.205$-- is partially corrected in our sample by adding in the OBA sample.  

Finally, as shown in Table \ref{tab.data}, the $[Fe/H]$ metallicity pa\-ra\-me\-ter is only available for the FGK-type stars in \citet{nordstrom04}. The metallicity calibration by \citet{holmberg07} provided new $[Fe/H]$ estimates for this sample. Differences between old and new estimates reach values up to $\pm0.2\dex$. \citep[see Fig. 7 in ][]{holmberg07}. However, \citet{reid07} compared the photometric estimate used in \citet{nordstrom04} with values obtained using echelle (high resolution) spectroscopy and found a dispersion of only $\sim$ 0.1 $\dex$. In addition, as shown by \citet{haywood06}, explicit biasing of the atmospheric parameters can lead to structures and spurious patterns in the age--metallicity diagrams. Undoubtedly, these considerations should be taken into account in the metallicity study of moving groups (Sect. \ref{feh}).

\section{Multiscale methods}\label{method}

The statistical methods that we use to characterize the structures (i.e. establish their shape, size or statistical significance) in the $U$--$V$--$age$--$[Fe/H]$ space are:

\begin{itemize}
\item Wavelet transform (WT): visualizing and detecting structures according to their different sizes or scales (Sect. \ref{wt}).

\item Wavelet denoising (WD): obtaining a smooth distribution function from a point distribution via a smoothing/filtering treatment at different scales that eliminates Poisson fluctuations (Sect. \ref{denoising}). 
\end{itemize}

A detailed description of these methods can be found in \citet{starck94,lega95,murtagh95,starck98,starck02}. In this section, we provide a brief description of the two-dimensional case. In later sections these methods are applied to different combinations of any two of the variables considered: kinematic parameters $U$ and $V$, age and metallicity.
To perform the calculations we use the MR software\footnote{http://thames.cs.rhul.ac.uk/$\sim$multires/} developed by CEA (Saclay, France) and the Nice Observatory. It consists of a set of tools that implements multi-scale methods for processing 1D signals, 2D images, and 3D data volumes.

\subsection{The wavelet transform}\label{wt}

\begin{table*}
\centering
\caption{Value of $\Delta$ and size of the structures detected at each smooth plane ($c_j$) and at each scale of the WT ($w_j$) for velocities, age and metallicity.}
\begin{tabular}{lr|lr|lr|lr|lr|lr}\hline\hline
\multicolumn{4}{c|}{$U$,$V$,$W$}&\multicolumn{4}{c|}{$age$}&\multicolumn{4}{c}{$[Fe/H]$}\\
\multicolumn{4}{c|}{($\kms$)}&\multicolumn{4}{c|}{($\Gyr$)}&\multicolumn{4}{c}{($\dex$)}\\\hline
\multicolumn{4}{c|}{$\Delta=0.5$}&\multicolumn{4}{c|}{$\Delta=0.01$}&\multicolumn{4}{c}{$\Delta=0.01$}\\\hline
$c_0$&0.5&&&$c_0$&0.01&&&$c_0$&0.01&&\\
$c_1$&1&$w_1$&0.75&$c_1$&0.02&$w_1$&0.015&$c_1$&0.02&$w_1$&0.015\\
$c_2$&2&$w_2$&1.5&$c_2$&0.04&$w_2$&0.030&$c_2$&0.04&$w_2$&0.030\\
$c_3$&4&$w_3$&3&$c_3$&0.08&$w_3$&0.060&$c_3$&0.08&$w_3$&0.060\\
$c_4$&8&$w_4$&6&$c_4$&0.16&$w_4$&0.120&$c_4$&0.16&$w_4$&0.120\\
$c_5$&16&$w_5$&12&$c_5$&0.32&$w_5$&0.240&$c_5$&0.32&$w_5$&0.240\\
$c_6$&32&$w_6$&24&$c_6$&0.64&$w_6$&0.480&$c_5$&0.64&$w_5$&0.480\\\hline
\end{tabular}
   \label{size}
\end{table*}

The WT decomposes a function $f(x,y)$ on the basis obtained by translation and dilation of the so-called  {\it mother wavelet}, which is localized in both physical and frequency space. The method consists of applying the correlation product between the function and the wavelet function:
\begin{eqnarray}
w_a(x,y) = f(x,y) \otimes \psi\left(\frac{x}{a}, \frac{y}{a}\right)
\end{eqnarray}
where $a$ is the scale parameter and $w_a(x,y)$ the wavelet coefficient. By varying $a$, we obtain a set of images, each of which corresponds to the wavelet coefficients of the data at a given scale. The integral of the wavelet function is equal to zero. Therefore, the WT analyses the overdensities and underdensities of the function, assigning them positive and negative coefficients respectively. A constant function would produce null coefficients. The key point is that the WT is able to discriminate structures as a function of scale, and thus it is well suited to detecting structures at one scale that are embedded within features at a different, larger scale.

It is possible to define a discrete WT which allows us to compute a discrete set of wavelet coefficients or a scale-related set of ``views'' of the 2-D function: the {\em \`a trous} WT algorithm. In this case, the WT performs on a grid of pixels with a bin size of $\Delta$. A set of band-pass filters that allows structures to be recognized at each scale is constructed from the so-called scaling function. The signal $c_0(x,y)$ can then be decomposed into a set $(w_1,..., w_J, c_J)$:
\begin{equation}\label{decomp}
   c_0(x,y)=\sum _{j=1}^{J}w_j(x,y)+c_J(x,y)
\end{equation}
where $c_J(x,y)$ is a smooth version of the original signal $c_0(x,y)$ and shows the details of $c_0(x,y)$ at scale of size $2^J$, in units of the grid bin, as they have been built with a smoothing filter of $2^J\times\Delta$ size. Equivalently, $c_j(x,y)$ is obtained with a smoothing filter of $2^j\times\Delta$ size. As $w_{j+1}=c_j-c_{j+1}$, the structures detected at each scale $j$ have a size that is approximately between $2^j\times\Delta$ and $2^{j+1}\times\Delta$. For more details of the {\em \`a trous} WT algorithm see e.g. \citet{starck02}.

In the case of a set of discrete points, the function $f(x,y)$ is first approximated by $c_0(x,y)$, which is obtained by smoothing the set of points on a grid with a bin size of $\Delta$. In our case this is obtained from simple star counts. The choice of $\Delta$ depends on observational errors or the resolution of the data. We choose 0.5 $\kms$, $0.01$ $\Gyr$ and $0.01$ $\dex$ for velocity, age  and metallicity, respectively. Table \ref{size} shows the size of the structures detected on each smooth plane and on each plane of the WT.

Here we use the {\em \`a trous} algorithm for the WT with the B$_3$-Spline as the scaling function. This leads to a row-by-row convolution with the mask $(1/16, 1/4, 3/8, 1/4, 1/16)$, followed by column-by-column convolution. This algorithm has several advantages. Due to the properties of the B$_3$-Spline, the transformation is isotropic and thus allows the detection of features with no preferred direction\footnote{Note that isotropy is not defined in non-metric spaces, such as some of the ones used here.}. Secondly, it enables us to restore the signal $c_0$ from the set of $w$ (Eq. \ref{decomp}) easily and without losing any information. Finally, the complexity of the computation of $c_0(x,y)$ of the initial grid is proportional to the number of points (stars) in the sample, whereas for the WT it is O($NJ$), where $N$ is the total number of pixels in the grid and $J$ is the number of scales. This results in a very low computational cost\footnote{The computing times are 4 s for the grid construction process with the whole sample (24190 stars) and 1 s for the WT in a typical grid of $400\times400$ pixels and $J=5$ with an Intel(R) Xeon(TM) processor at 3.0GHz.}.

\subsection{The denoising method}\label{denoising}

The data $c_0(x,y)$ present statistical fluctuations related to the fact that the sample is finite. The noise of the signal (number of stars) in each initial grid bin obeys Poisson statistics. These fluctuations are detected as structures, especially at the smaller scales of the WT. The aim of WD is to ascertain the significance of the wavelet coefficients $w_j(x,y)$ at each scale, which leads to adaptative filtering.

If a model for the noise can be assumed, the probability that a wavelet coefficient $w_j(x,y)$ is significant can be estimated. Then in simple thresholding methods, a detection threshold is defined for each scale and coefficients with higher probability of being due to noise are rejected. But as an alternative, we use Wiener-like filtering in the wavelet space, or multiresolution Wiener filtering \citep{starck94}. This allows the treatment of each coefficient significance as a continuous function and thus denoising consists of weighting each coefficient according to its significance. This filtering method is detailed in Sect. \ref{wiener}.

The WD method allows us to obtain a smooth distribution function by reconstructing the transformed data after denoising or filtering at each scale, which is the key point of the method. Thus, the denoised signal $\widetilde{c}(x,y)$ is obtained by adding the first  $J$ denoised scales $\widetilde{w_j}$ ($j=1, J$) to the smooth version  $c_J(x,y)$:

\begin{equation}\label{comp}
 \widetilde{c}(x,y)=\sum _{j=1}^{J}\widetilde{w_j}(x,y)+c_J(x,y)\ .
\end{equation}

The number of scales\footnote{Note that the parameter $n$ in the MR software is $n=J+1$.} $J$ which should be used in the WD depends on the image or signal size. In the MR software documentation it is stated that, in theory, this could be $J=\log_2(N)$, where $N$ is the number of pixels of the grid in its smallest direction, but it is also suggested that, in practice, it is preferable to use a lower value such as $J=\log_2(N)-1$ or $J=\log_2(N)-2$. It seems reasonable to adopt the option of increasing $J$ until no change is observed in the reconstructed denoised signal $\widetilde{c}(x,y)$, i.e. WD up to the scale $J$, hereafter called $J_{plateau}$, where all signal is found to be significant (noise free). This has been the option followed in most of the analyses of this study. Actually, in most cases $J_{plateau}$ is equal to or even lower than $\log_2(N)-1$ or $\log_2(N)-2$. Nevertheless, for some specific cases, especially when dealing with a signal with fewer counts per pixel, we find that $J_{plateau}$ is higher than the practical values proposed.
Most probably, in these cases, WD up to $J_{plateau}$ could imply very few data values in the final scales and consequently, too much loss of signal. These cases will be clearly flagged, as we are aware that conclusions derived from them would demand the use of larger samples for definitive confirmation. In these situations the $J$ up to which the WD has been carried out will be specified, whereas in all other cases WD has been done up to $J_{plateau}$.

WD offers several advantages since the multiscale structures that we expect to appear could be very complex. The method is more straightforward than other methods used in this field. It offers smoothing with a unique recipe and it does not require additional simulations for the treatment of Poisson fluctuations. Furthermore, it allows an automatic local filtering as the denoising is carried out at several scales. The analysis is not restricted to one specific scale or band-width because all scales are visualized at the same time in the final distribution. Finally, this method is more precise than other smoothing methods such as Gaussian smoothing which degrades resolution and is shown to introduce Gaussian features into the distribution \citep{martinez05} and, therefore, is not suitable for the detection of structures that may be far from Gaussian.

\subsubsection{The multiresolution Wiener filtering}\label{wiener}

In multiresolution Wiener filtering \citep{starck94}, it is assumed that a measured wavelet coefficient $w_j$, at a given scale $j$ and a given position, results from a noisy process, with a Gaussian distribution with a mathematical expectation $W_j$, and a standard deviation $\sigma_j$:

\begin{equation}\label{p1}
\displaystyle P(w_j/W_j) = \frac{1}{\sqrt{2\pi}\sigma_j} e^{- \frac{(w_j-W_j)^2} {2\sigma_j^2}}\ .	  	
\end{equation}

If the noise in the data $c_0$ is Poisson noise, as in the present study, the Anscombe transformation can be applied to turn it into a stationary Gaussian noise with unitary variance under the assumption that the mean value of $c_0$ is large \citep{anscombe48}. We expect the counts associated with the structures in our study to be large enough for this condition to hold. After the Anscombe transformation is performed, the probability density of the coefficients  $w_j$ becomes Gaussian and the multiresolution Wiener filtering can be applied. The noise standard deviation at each scale $\sigma_j$ resulting from a Gaussian noise with standard deviation equal to 1 are calculated in \citet[][]{starck02} (see their Table 2). Then multiresolution Wiener filtering takes into account that the set of expected coefficients $W_j$ for a given scale also follows a Gaussian distribution, with a null mean, as the wavelet function has null integral, and a standard deviation $S_j$:

\begin{equation}\label{p2}
\displaystyle P(W_j) = \frac{1}{\sqrt{2\pi}S_j}e^{-\frac{W_j^2}{2S_j^2}} \ .	  \end{equation}

In the algorithm, $S_j$ is locally estimated as $S_j^2 = s_j^2 - \sigma_j^2$, where $s_j^2$ is the variance of $w_j$. To estimate $W_j$ knowing $w_j$, Bayes' theorem gives:
\begin{equation}\label{p4}
 P(W_j/w_j) = \frac{P(W_j)P(w_j/W_j)}{P(w_j)}=\frac{1}{\sqrt{2\pi}\beta_j}e^{-\frac{(W_j-\alpha_j w_j)^2}{2\beta_j^2}}  	  
\end{equation}
\noindent where $\alpha_j = \frac{S_j^2}{S_j^2+\sigma_j^2}$ and $\beta_j^2 = \frac{S_j^2\sigma_j^2}{S_j^2 + \sigma_j^2}$. Thus, the probability $P(W_j/w_j)$ follows a Gaussian distribution with a mean $\alpha_j w_j$ and a variance $\beta_j^2 $. The mathematical expectation of $W_j$ is $\alpha_j w_j$ and consequently the denoised coefficients are computed through a linear filter with a simple multiplication of the coefficients $\widetilde{w_j}=\alpha_jw_j$. The complexity of the WD using this filtering method is O($NJ$), where $N$ is the total number of pixels of the grid and $J$ is the number of scales\footnote{The computing time for a typical grid of $400\times400$ pixels and $J=5$ is 4 s with an Intel(R) Xeon(TM) processor at 3.0GHz.}. 

This and other algorithms for filtering in the wavelet space were compared in \citet{starck94}. Using an image with artificially added noise, they conclude that Wiener-type filtering provided the best results. Although it is out of the scope of the present study to investigate exhaustively the different filtering methods, we have compared the results obtained using a thresholding method with the multiresolution Wiener filtering. We conclude that the latter produces smoother distributions, without additional artifacts.

\begin{figure*}
    \centering
    \includegraphics[width=0.7\textwidth]{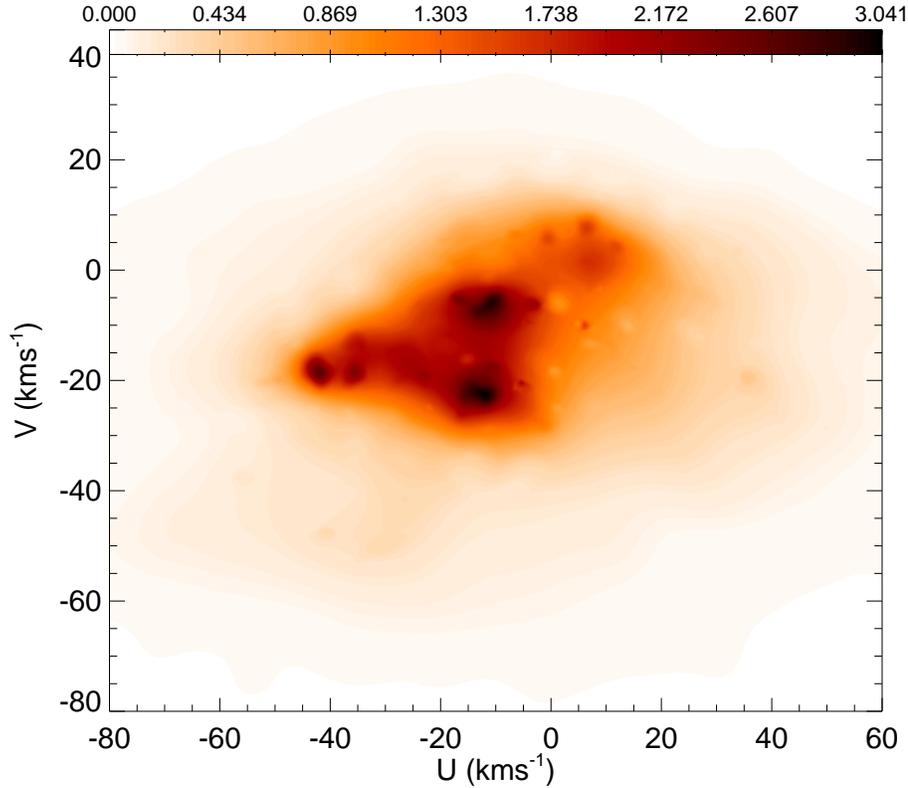}
    \caption{Density field in the $U$--$V$ plane for the whole observational sample obtained by WD ($J_{plateau}=4$).}
    \label{UV123456}
\end{figure*}

\section{Structures in velocity space}\label{velocity}

\begin{figure}
    \centering
    \includegraphics[width=0.49\textwidth]{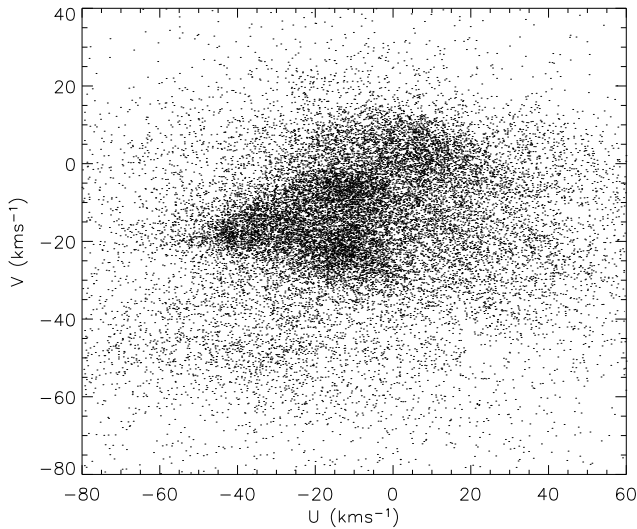}
 \caption{Stellar velocities in the $U$--$V$ plane for the whole observational sample.}
    \label{points}
\end{figure}

Figure \ref{UV123456} shows the velocity distribution in the $U$--$V$ plane\footnote{We deal only with these two components of the velocity because classic moving groups are less visible in the $W$ distribution due to the more thorough phase-mixing of the movement perpendicular to the Galactic plane \citep{dehnen98,seabroke07}.} for the whole sample of 24910 stars obtained by means of the WD method. This figure can be compared to Fig. \ref{points} where the velocities of the individual stars are plotted as dots. Table \ref{t_groups} and Fig. \ref{groups_branches} show the positions of the centres of the classic moving groups according to \citet{dehnen98} and \citet{eggen71,eggen96}\footnote{The list could be far more extensive; only the most significant groups have been chosen. Similar values for their components have been reported by other authors \citep[see e.g.][]{montes01,lopezsantiago06}.}. Connections and continuity can be seen between these classic kinematic structures. As was first pointed out by \citet{skuljan99}, these connections arrange classic moving groups in several branches whose approximate positions can be shown also in Fig. \ref{groups_branches}. These superstructures (the branches), structures (moving groups) and their substructures are characterized in this section.

\subsection{Classic moving groups}\label{classic}

\begin{table}
\caption{Heliocentric velocities of the main moving groups according to (1) \citet{dehnen98} and (2) \citet[ and references therein]{eggen71,eggen96}.}
\label{t_groups}
\centering
\begin{tabular}{l|rr|rr} \hline\hline
                      &\multicolumn{2}{c|}{(1)}& \multicolumn{2}{c}{(2)}\\
\small Moving group&\multicolumn{1}{c}{$U$}&\multicolumn{1}{c|}{$V$}&\multicolumn{1}{c}{$U$}&\multicolumn{1}{c}{$V$}\\
& \scriptsize {($\kms$)}&\scriptsize {($\kms$)}&\scriptsize {($\kms$)}&\scriptsize {($\kms$)}\\\hline
1 Pleiades (Stream 0) &$-12$   &$-22$    &$-11.6$ &$-20.7$\\
2 Hyades (Stream I)   &$-40$   &$-20$    &$-40.4$ &$-16.0$\\
3 Sirius (Stream II)  &$ 9 $   & $ 3 $   &$14.9$  &$1.4$\\
4 Coma Berenices      &$-10$   & $-5$    &      &\\
5 NGC 1901            &$-25$   &$-10$    &$-26.4$ &$-10.4$\\
6 HR1614              &$ 15$   &$-60$    &$5.8$   &$-59.6$\\
7                     &$ 20$   &$-20$    &      &\\
8                     &$-40$   &$-50$    &      &\\
9                     &$-25$   &$-50$    &      &\\
10                    &$ 50$   &$  0$    &      &\\
11                    &$ 50$   &$-25 $   &      &\\
12 IC 2391            &      &       &$-20.8$ &$-15.9$\\
13                    &$-70$   &$-10$    &      &\\
14                    &$-70$   &$-50$    &      &\\
15 61 Cygni           &      &       &$-80$   &$-53$\\
16 $\zeta$ Herculis   &      &       &$-30$   &$-50$\\
17 Wolf 630           &      &       &$25 $   &$-33$\\\hline
\end{tabular}
\end{table}

\begin{figure}
    \centering
  \includegraphics[width=0.4\textwidth]{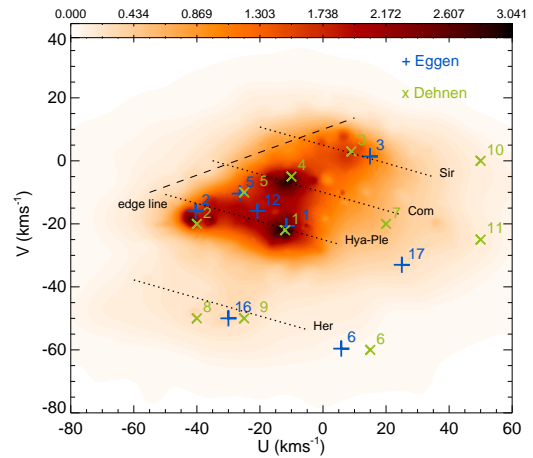}
 \caption{As Fig. \ref{UV123456} but with the superposition of the classic moving groups in Table \ref{t_groups}, the approximate trace of the branches and the trace of the edge line (see text in Sect. \ref{branches}).}
    \label{groups_branches}
\end{figure}

The established understanding of classic moving groups immediately leads to the following points when analysing Fig. \ref{groups_branches}:

\begin{itemize}
\item The Sirius, Coma Berenices, Hyades and Pleiades moving groups are clearly identified but they have neither a clearly defined shape nor defined limits. Furthermore, they present substructure.

\item The centre of the Sirius moving group is not well-defined, which could explain the discrepancies in its position on the $U$--$V$ plane found in the literature. This moving group seems to be better described by a branch-like shape with a clear extension.

\item Similar arguments can be applied to the Hercules moving group, although it is less prominent in Fig. \ref{groups_branches} due to its low density. The two peaks proposed by \citet{dehnen98} (groups 8 and 9 in Table \ref{t_groups}) are faintly observed but this structure appears to be a continuous elongated feature.

\item The Hyades and Pleiades moving groups are clearly grouped together, forming just one branch. The centres established by Eggen and Dehnen for NGC 1901 and IC 2391 (groups 5 and 12 in Table \ref{t_groups}) seem to be placed slightly above the crest of the branch (less negative $V$), in a low density region within the distribution.

\item  Groups such as NGC 1901 (group 5), HR 1614 (group 6) or IC 2391 (group 12) are not observed, possibly due to the small contribution they make compared to the main structures. 

\item New structures such as that centred at $(35,-20) \kms$, which is weak and does not have a well-defined shape, may all be considered as part of the elongation of the Sirius or Coma Berenices structures, rather than being necessarily linked to the nearest identified groups (11 or 17 in this case).

\end{itemize}

\begin{figure}
    \centering
 \includegraphics[width=0.35\textwidth]{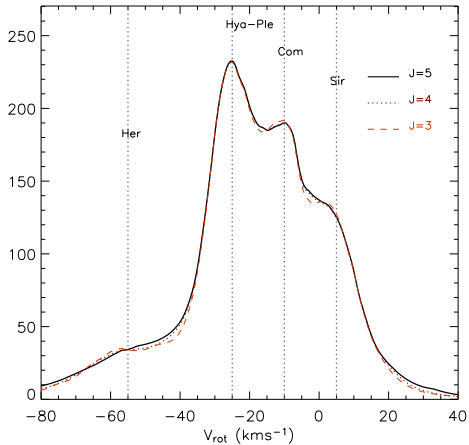}
 \caption{Density field for the $V_{rot}$ component obtained integrating the density in the $U_{rot}$--$V_{rot}$ plane. The approximate positions of the four branches of Hercules, Hyades-Pleiades, Coma Berenices and Sirius are marked with vertical lines according to the maximums.}
    \label{V_branches}
\end{figure}

\subsection{Kinematic branches}\label{branches}

\citet{skuljan99} detected the existence of at least three long, parallel and equidistant branches in the $U$--$V$ plane: the {\it Sirius branch}, the {\it middle branch} (here Coma Berenices branch) and the {\it Pleiades branch} (here Hyades-Pleiades branch). The increased extent of our sample, in terms of both luminosity and spectral type, allows us to characterize these three branches and also the Hercules branch. In this section the properties of these branches in the whole sample are studied and Sect. \ref{spectral} deals with the subsamples of different spectral type.

To simplify the study of the branches, a clockwise rotation through an angle $\beta$ is applied to the original $(U,V)$ components. In the new coordinate system $(U_{rot},V_{rot})$ the branches are better aligned with the horizontal axis (except for their slight curvature). Although \citet{skuljan99} proposed a turn of $ 25\deg$, a value of $\beta \sim 16\deg$ is more suitable for the data we present here, especially for the three main branches. Their approximate positions are delineated as dotted lines with this slope in Fig. \ref{groups_branches}.

Figure \ref{V_branches} shows the density along the $V_{rot}$ component for the whole sample obtained integrating the density in the $U_{rot}$--$V_{rot}$ plane. The WD has been carried out up to different $J$ (3, 4, 5) in order to illustrate the discussion regarding the choice of $J$ in Sect. \ref{denoising}. The four branches are clearly seen and their approximate positions are marked with vertical lines. It can be noticed that $J$ has become rapidly $J_{plateau}$ and results almost do not change from $J$ to $J$. Our data confirm the nearly equidistant character of the branches. The first three branches are found to be separated in intervals of approximately 15 $\kms$. The possible dependence of this interval on spectral type or age is discussed in Sects. \ref{spectral} and \ref{age}. The separation between the Hyades-Pleiades and the Hercules branch is $\sim 30\kms$.

\begin{figure}
   	\centering 
	\includegraphics[width=0.35\textwidth]{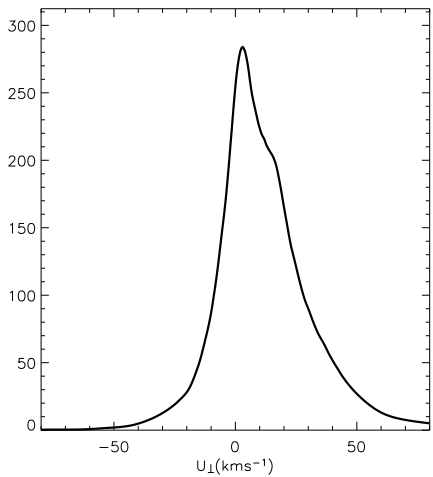}
	\caption{Density field in the direction perpendicular to the edge line ($U_{\perp}$) obtained integrating the density in the $U$--$V$ plane for the whole observational sample ($J_{plateau}=4$).}        
  	\label{U_edge}
\end{figure}

\begin{figure}
   	\centering 
	\includegraphics[width=0.35\textwidth]{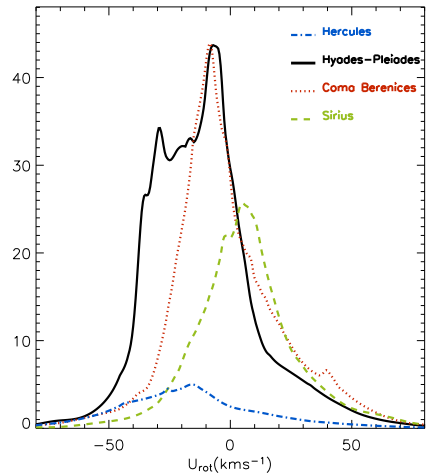}
	\caption{Density field for the $U_{rot}$ component obtained integrating the density in the $U_{rot}$--$V_{rot}$ plane for each of the four branches ($J_{plateau}=4$).}        
  	\label{U_branches}
\end{figure}

\citet{skuljan99} noticed the existence of a fairly sharp 'edge line' at $30 \deg$ to the $U$ axis and connecting the extremes of the branches at low $U$ and high $V$ (see their Fig. 5). This feature can be evaluated by computing the density field in the direction perpendicular to this hypothetical 'edge line' (direction of $U_{\perp}$ from now on). First, several directions were scanned to look for the sharpest distribution: a lower inclination of about $20 \deg$ was found here. This edge line is plotted in Fig. \ref{groups_branches} and the perpendicular density distribution along $U_{\perp}$ is presented in Fig. \ref{U_edge}. Undoubtedly, the asymmetry and the long tail at positive $U_{\perp}>0$ observed would support the existence of a sharp edge line. However, we want to stress that: 
\begin{itemize}
\item The upper part of the edge line is not empty: a considerable number of stars occupy this region of the $U$--$V$ plane (see Fig. \ref{points}).
\item Equivalent importance should be given to other similar 'edge lines' such as the sharp distribution observed in Figs. \ref{points} and \ref{UV123456} near the Hyades-Pleiades branch towards low $U$ and low $V$, which is shown in Fig. \ref{V_branches} with the drop at the left of this branch.
\item Each branch presents a different density distribution near the edge line (Fig. \ref{UV123456}). This is also confirmed by examining the density characteristics along each branch, i.e. in the direction of the component $U_{rot}$, plotted in Fig. \ref{U_branches}. These density distributions are obtained from the whole $U_{rot}$--$V_{rot}$ distribution considering the central positions in $V_{rot}$ of the branches according to the maximums in Fig. \ref{V_branches} ($-55$, $-25$, $-10$ and $5$ $\kms$ for the Hercules, Hyades-Pleiades, Coma Berenices and Sirius branches, respectively) and a width of $\pm 4\kms$. It can be seen that the slope of the density drop at negative $U_{rot}$ is different for each branch, being considerably steeper for the Hyades-Pleiades branch in part due to the gathering of stars of the small substructures at the very beginning of the branch. The drop becomes less abrupt from Coma Berenices to Sirius to the Hercules branch. Moreover, the Hyades-Pleiades and Coma Berenices branches show a slightly asymmetric distribution, with a longer tail of stars at positive $U_{rot}$. 
\end{itemize}
The existence of all these abrupt features in the $U$--$V$ plane definitely rules out the classic idea of a smooth velocity field distribution and it even questions the scenario where several structures are superimposed on this smooth field.

\subsection{Spectral type analysis}\label{spectral}

\begin{figure*}
    \centering 
 \includegraphics[width=0.38\textwidth]{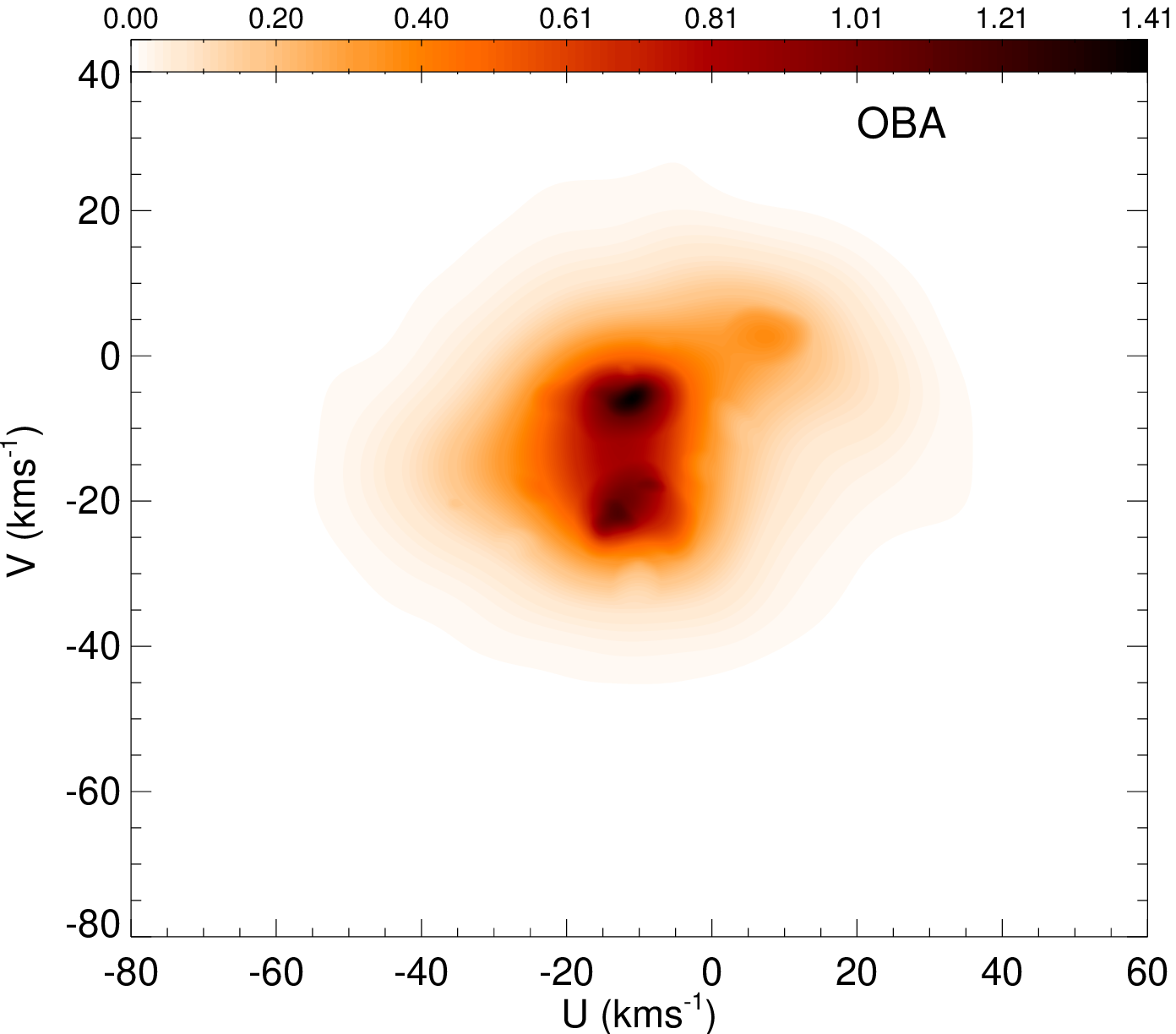}
 \includegraphics[width=0.33\textwidth]{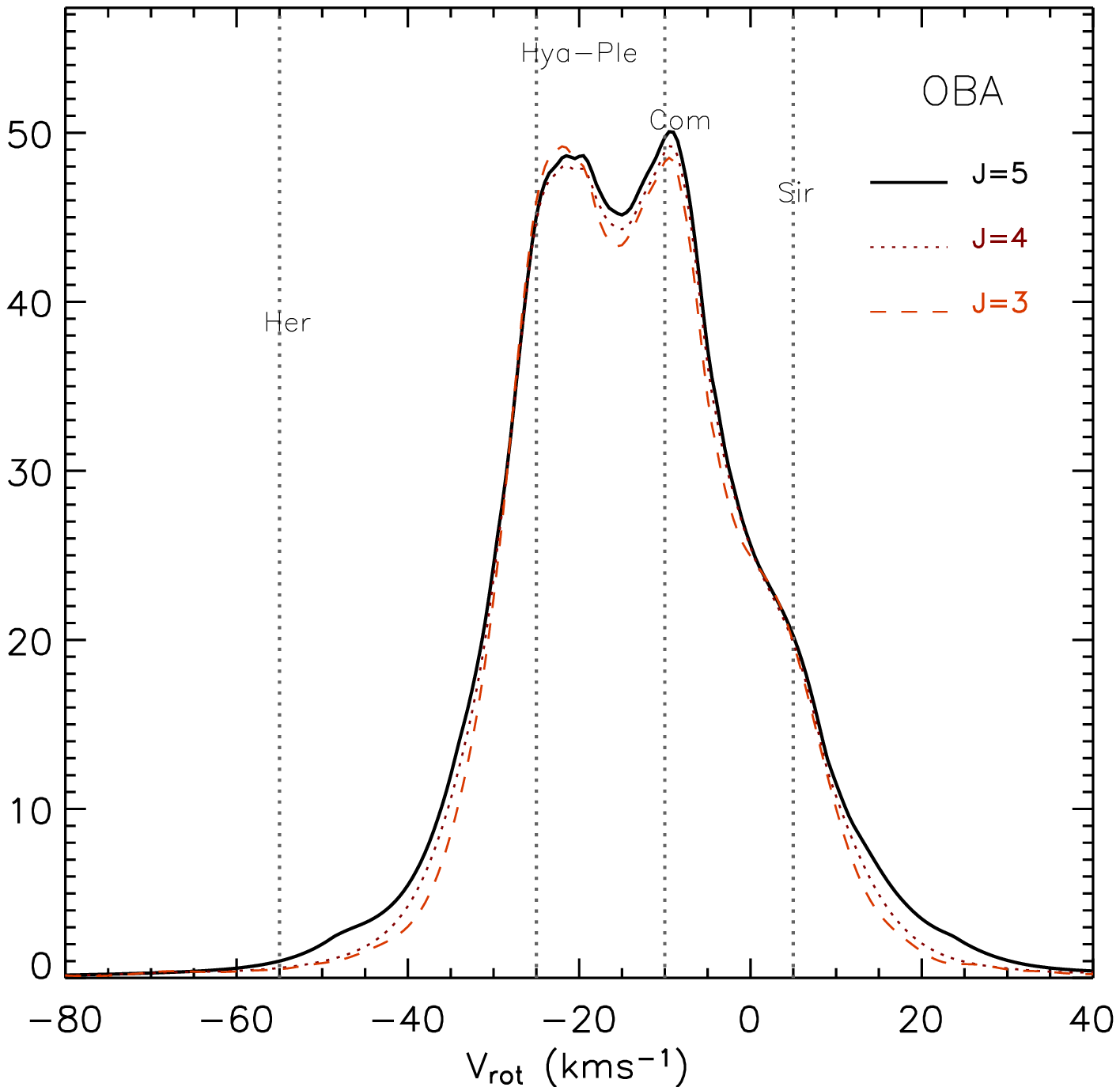}

 \includegraphics[width=0.38\textwidth]{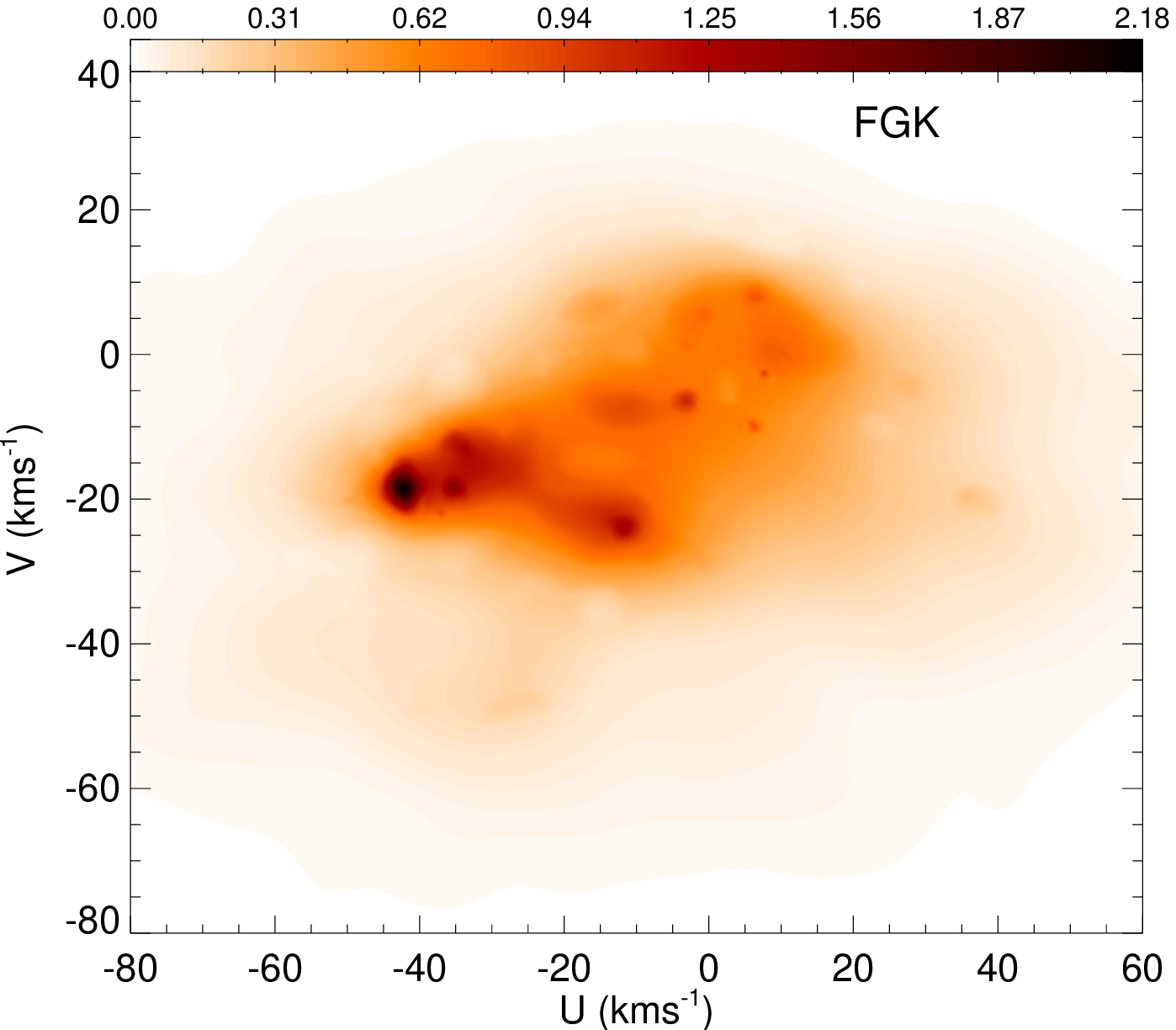}
 \includegraphics[width=0.33\textwidth]{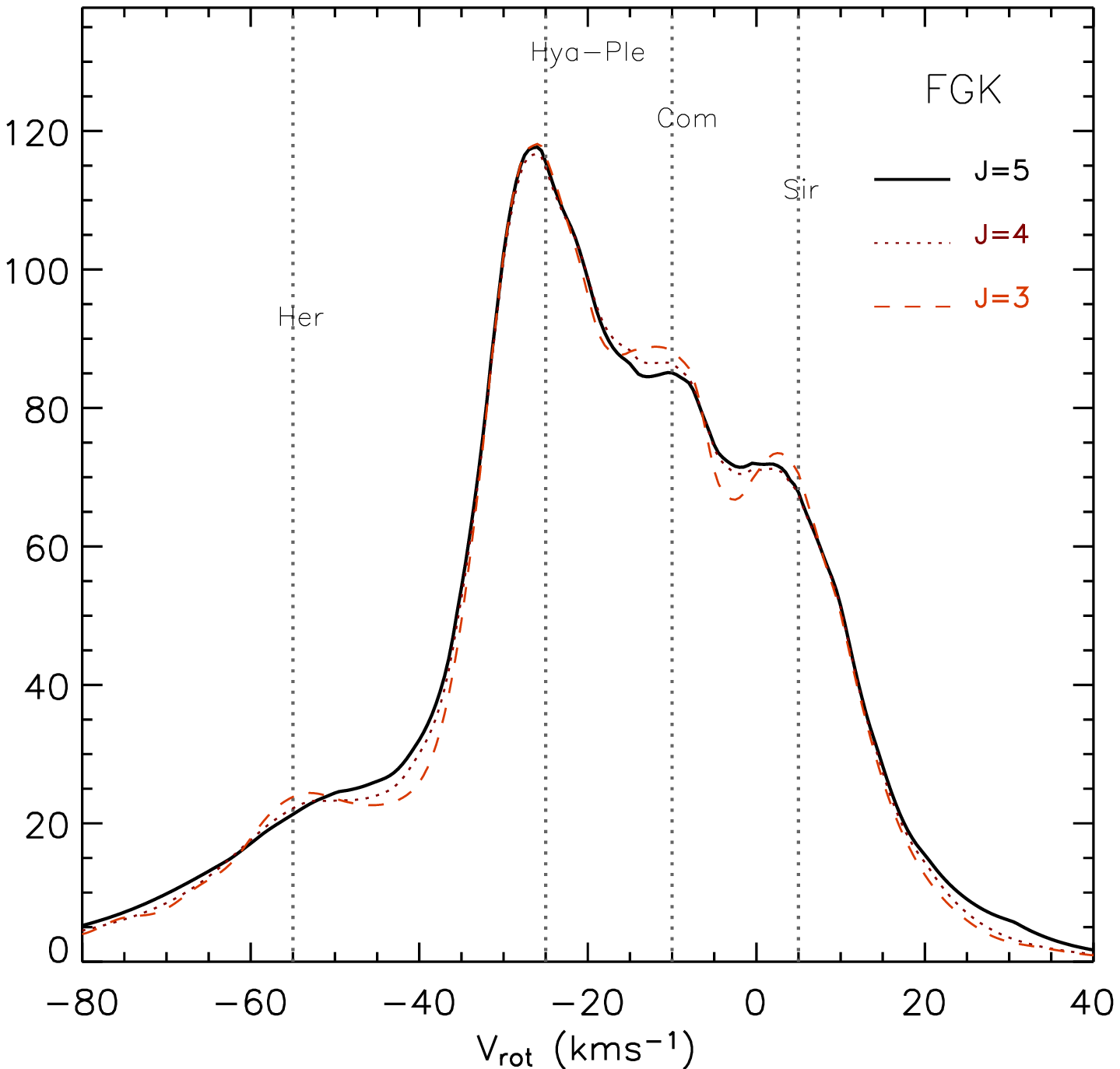}

 \includegraphics[width=0.38\textwidth]{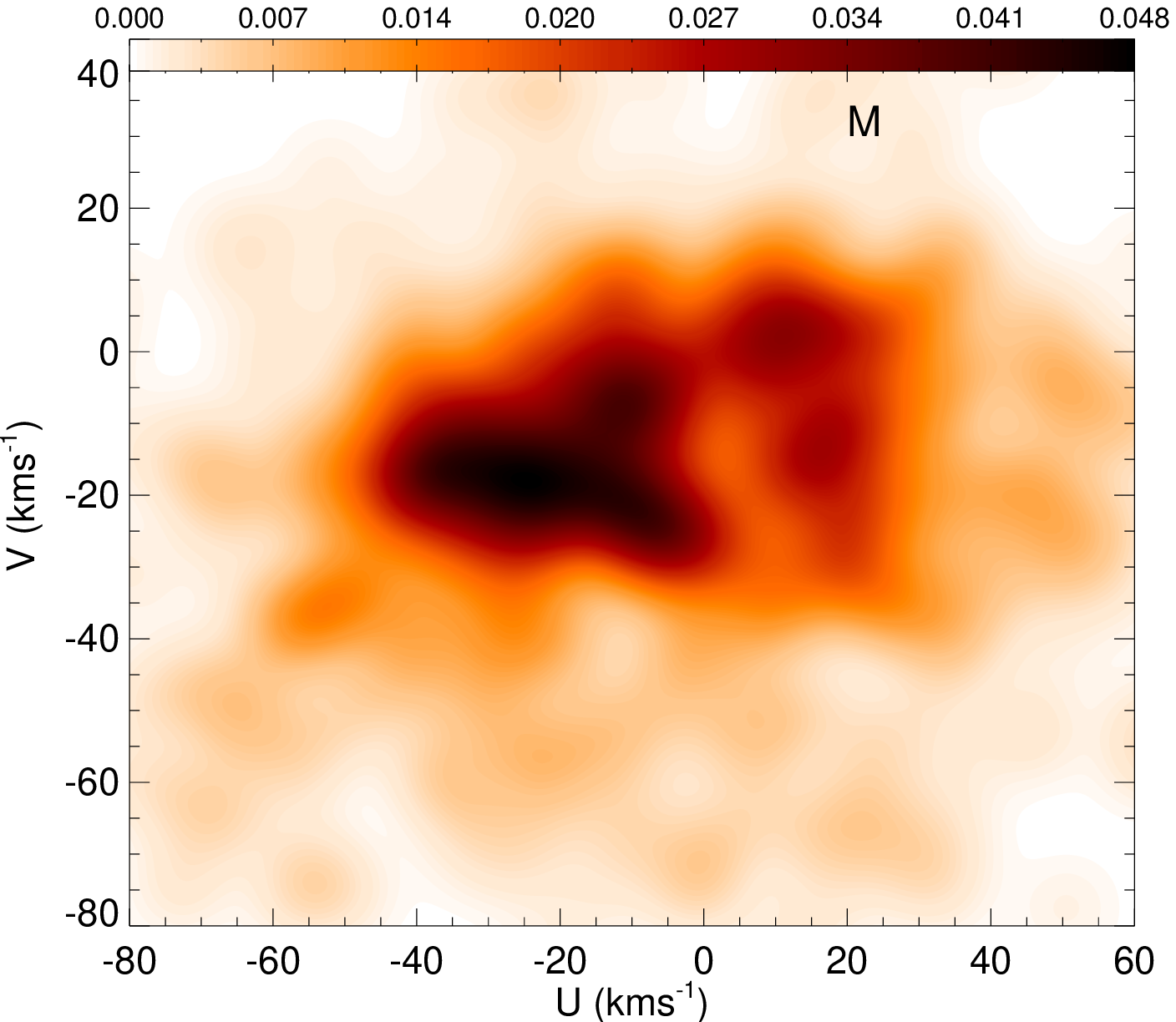}
 \includegraphics[width=0.33\textwidth]{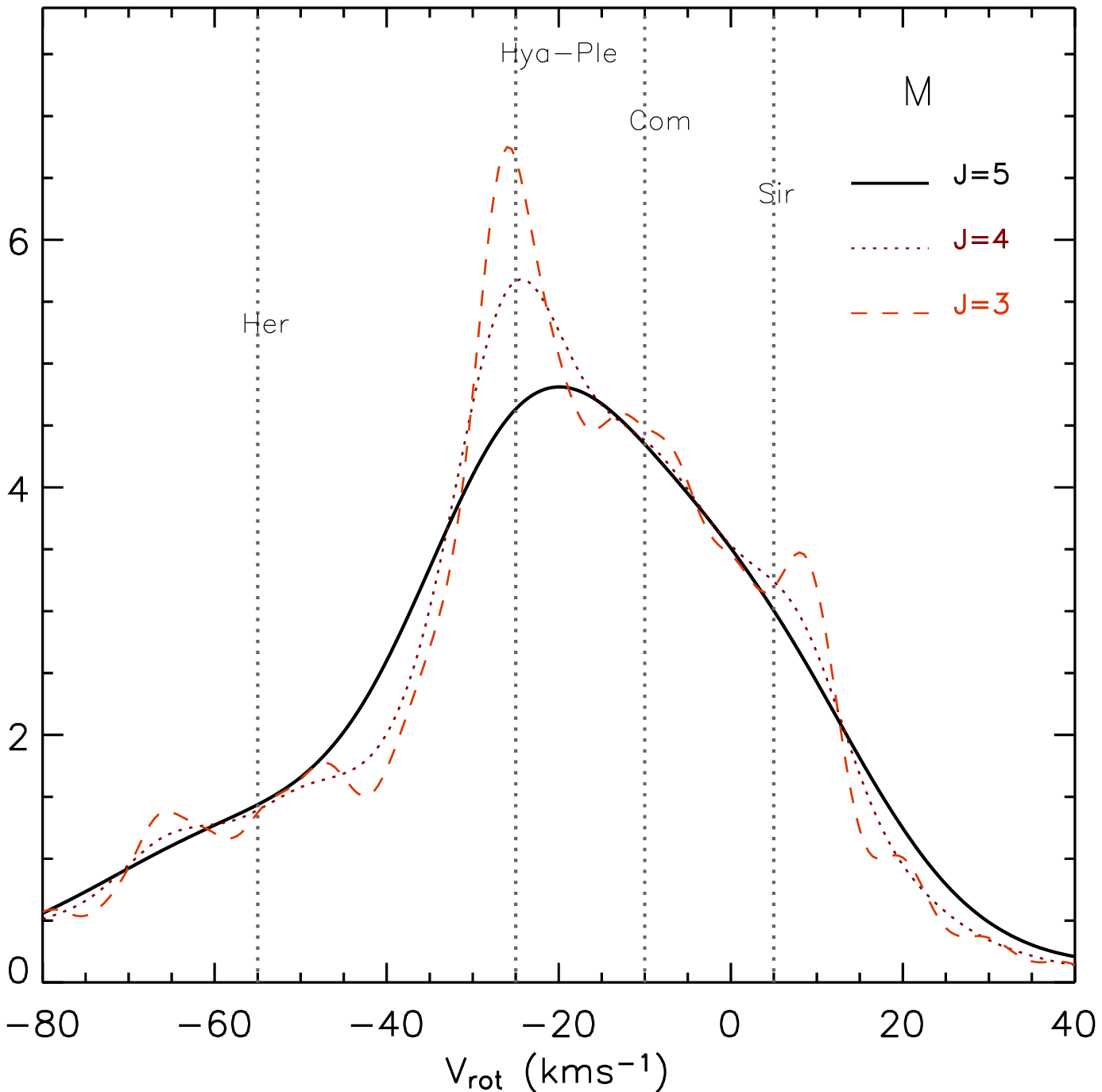}

 \includegraphics[width=0.38\textwidth]{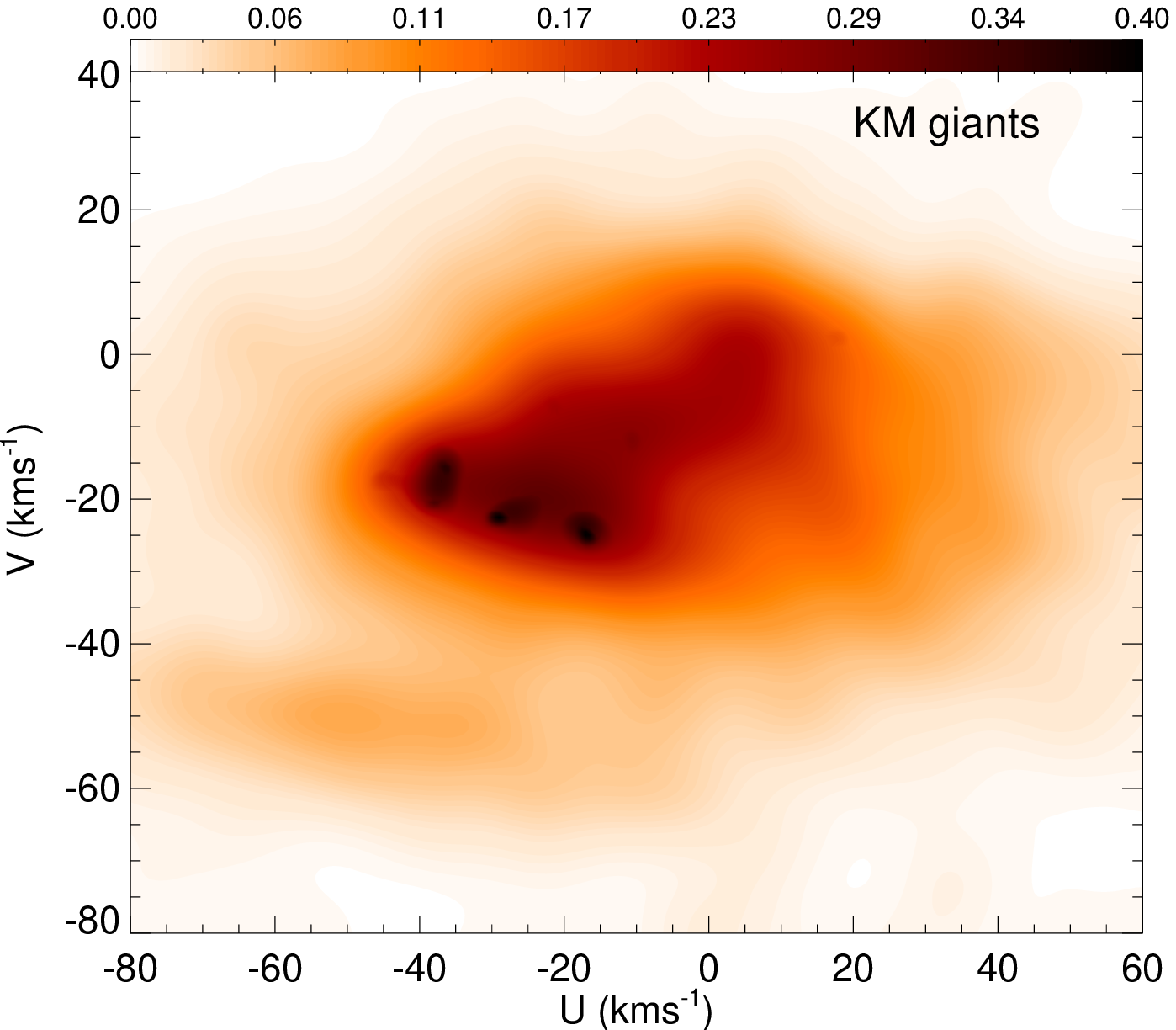}
 \includegraphics[width=0.33\textwidth]{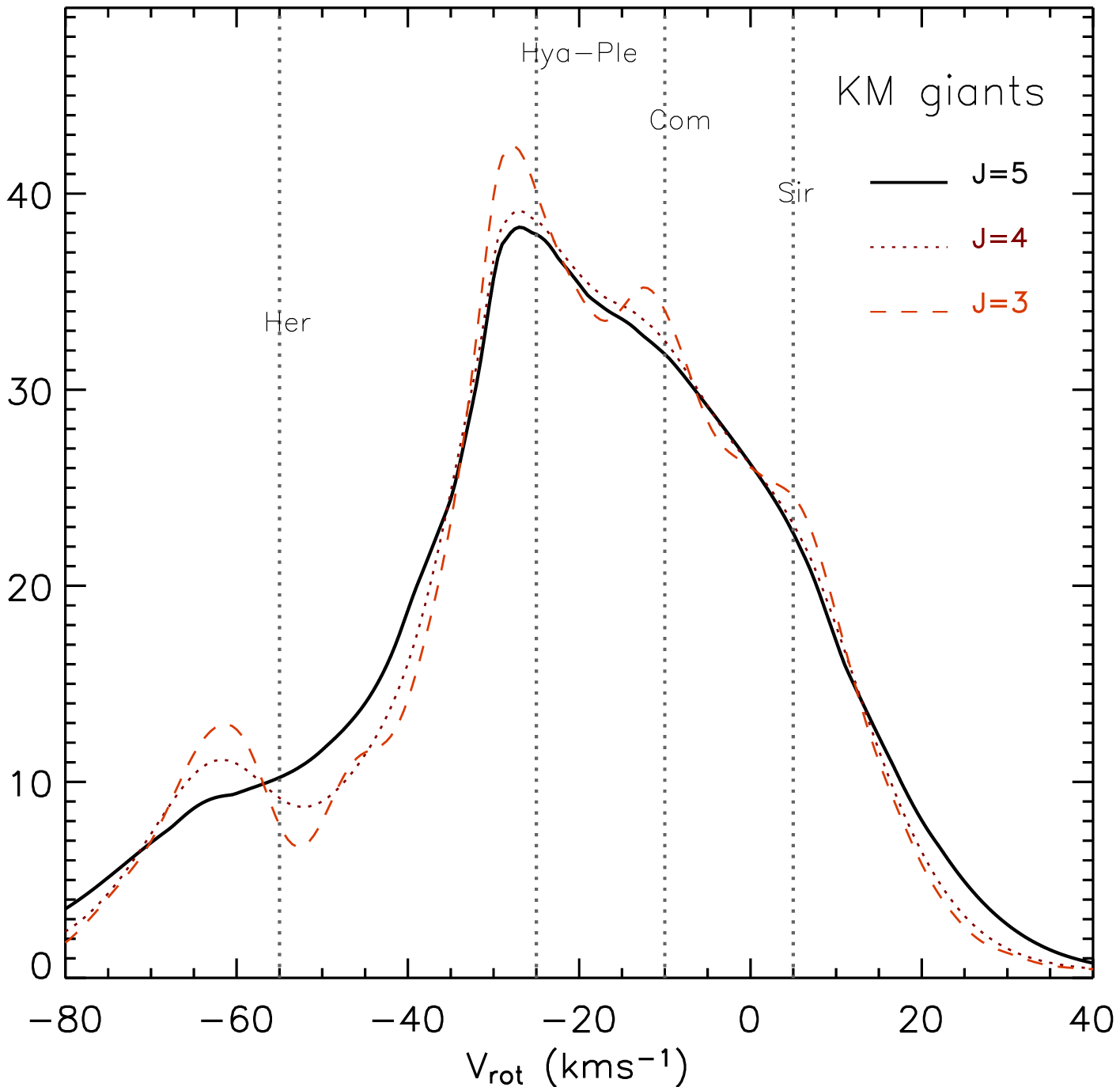}
    \caption{{\bf Left column:} Density field in the $U$--$V$ plane for the the subsamples of OBA, FGK, M dwarf stars and KM giant stars obtained by WD with $J=4$. {\bf Right column:} Density field for the $V_{rot}$ component obtained integrating the density in the $U_{rot}$--$V_{rot}$ plane and with different values of $J$ of the WD: 3, 4 and 5.}
    \label{spectraltypes}
\end{figure*}

Figure \ref{spectraltypes} shows the density field in the $U$--$V$ plane (left) and the density for the $V_{rot}$ component (right) for the subsamples of OBA, FGK, M dwarfs and KM giants. As mentioned in Sect. \ref{introduction}, some of these subsamples have been used for moving group studies by different authors and using different techniques. This is the first time that the same statistical method is applied to all of them. In this sense, a direct comparison among subsamples deserves special attention.

In the right column plots, the WD has been carried out up to different $J$ (3, 4, 5), as in Fig. \ref{V_branches} for the whole sample, in order to return to the considerations about the choice of $J$ in Sect. \ref{denoising}. The maximum value of $J$ presented in these plots is always higher than or equal to $J_{plateau}$, which is the one for which the result does not change when higher scales are denoised. For the FGK and OBA subsamples, $J$ becomes rapidly $J_{plateau}$ and the signals almost do not change from $J$ to $J$. Therefore, the significance of the structures and branches present in these distributions is strongly supported. However, for the subsamples of M dwarfs and KM giants the peaks whose existence has been independently proved by the previous robust results here are drastically diluted when denoising up to the $J_{plateau}$. For this and to avoid the loss of signal (Sect. \ref{denoising}), we present and discuss the $U$--$V$ distributions with $J=4$ for all subsamples despite this not being $J_{plateau}$ for all of them.

Some of the well-known characteristics according to spectral type are observed in these figures, for example the predominance of the Coma Berenices and Pleiades groups or the absence of the Hyades moving group for the OBA stars. As some of these features depend on age, they will be studied in Sect. \ref{age}. More importantly, the three branches of Hyades-Pleiades, Coma Berenices and Sirius are detected in all subsamples (Fig. \ref{spectraltypes}, right). \citet{skuljan99}, with a different rotation angle $\beta$, suggested a separation of 15$\kms$ for early-type stars but 20$\kms$ for late-type stars. Here, the value of 15$\kms$ is more suitable for all subsamples. In all cases, the positions of these three overdensities differ by less than $3$-$4\kms$ from the values derived for the whole sample (vertical lines).

The comparison between the velocity distributions of the KM giants and the FGK dwarfs deserves special attention since they have very different space volume coverage (see Fig. \ref{hisdist}) and the number of stars in both subsamples is comparable. A similar examination was carried out by \citet{seabroke07}. We agree with these authors that the kinematic structures are well maintained in both distributions and thus, considering their different spatial extension, moving groups cannot be simple remnants of star clusters. Our method allows us to go further in this comparison. For instance, a clear overdensity appears at $(U,V)\sim(-27,-22)\kms$ in the middle of the Hyades-Pleiades branch for the KM giants sample which is not present for the FGK dwarfs. Most specially, the Hercules branch presents significant differences both in shape and in position of the density maximum. Whereas for the KM giants this branch is seen as a clear elongated structure, for the FGK dwarfs there is a density maximum more localized in the $U$--$V$ plane. Moreover, Fig. \ref{spectraltypes} (right) shows an evident discrepancy between the positions of the Hercules branch located at  $V_{rot}\sim-62\kms$ for the KM giants but at $V_{rot}\sim-55\kms$ for the FGK dwarf stars. Two possible explanations for this discrepancy are: i) errors in distance estimates or biases and ii) real differences due to the different volume coverage or Galactic position. Previous discrepancies in the position of the Hercules group are seen by checking the literature. \citet{fux01} found Hercules centred at ($U,V$)=($-35,-45$)$\kms$ using a sample of nearby stars ($d<100\pc$), whereas a value of ($-42,-51$)$\kms$ was obtained by \citet{famaey05} for the KM giants. These values correspond to $V_{rot}=-53$ and $-61\kms$, respectively. Also, as mentioned in Sect. \ref{classic}, \citet{dehnen98} detected two peaks in the region of this branch (groups 8 and 9 in Table \ref{t_groups}) at $V_{rot}=-59$ and $-55\kms$. Although density values in the tail of the distribution where Hercules is placed are marginally significant, we will pay special attention to this discrepancy.

As mention in Sect. \ref{data}, no significant bias is expected for the FGK stars with trigonometric parallaxes (with small relative errors of less than 13\%). We have checked that the position of the Hercules branch does not change at all when separately considering the FGK stars with trigonometric parallax or those with photometric distances. Consequently, this establishes the existence of the peak at $V_{rot}\sim-55\kms$ for stars nearer than $100$-$150\pc$.

\begin{figure}
    \centering 
 \includegraphics[width=0.38\textwidth]{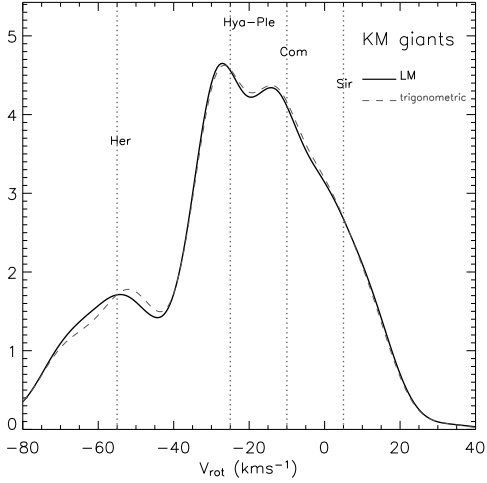}
    \caption{Density field for the $V_{rot}$ component obtained integrating the density in the $U_{rot}$--$V_{rot}$ plane for $J=4$ of the WD for the KM giants with $\sigma_\pi/\pi < 10$\% (756 stars) using the trigonometric distance (dashed line) and the LM distance (solid line).}
    \label{V_lmdist}
\end{figure}

More importantly, we have selected stars from the subsample of KM giants with relative errors in trigonometric parallaxes $\sigma_\pi/\pi<10$\%. As mentioned, biases in trigonometric distances are negligible for this truncation of the relative error.
In Fig. \ref{V_lmdist} we compare the density in the $V_{rot}$ component obtained calculating the velocity using: i) the distance estimates from the method in \citet{luri96} --LM distances from now on-- (solid line) and ii) the distances from trigonometric parallaxes (dashed line). As expected, both curves are practically identical using LM and trigonometric distances since high-quality information on trigonometric parallaxes was used in the derivation of the LM distances. Remarkably, we observe that the overdensity of the Hercules branch appears at $V_{rot}\sim-55\kms$, as for the FGK dwarfs. With this cut in $\sigma_\pi/\pi$, we are in fact selecting stars with distances $<150\pc$, which proves again the existence of a real kinematic structure at $V_{rot}\sim-55\kms$ for the nearer stars. Note that, by contrast, we find that the velocity distributions of the stars located in shells of LM distances further than 100$\pc$ centred in the Sun\footnote{Note that due to its source of radial velocities, this subsample is restricted to the northern equatorial hemisphere \citep[see ][]{famaey05} and therefore, stars in these shells are not isotropically distributed around the Sun.} do present always the peak at $\sim-62\kms$ using the LM distances (figures are omitted). Therefore, the situation is complex.

\begin{figure}
    \centering
 \includegraphics[width=0.26\textwidth]{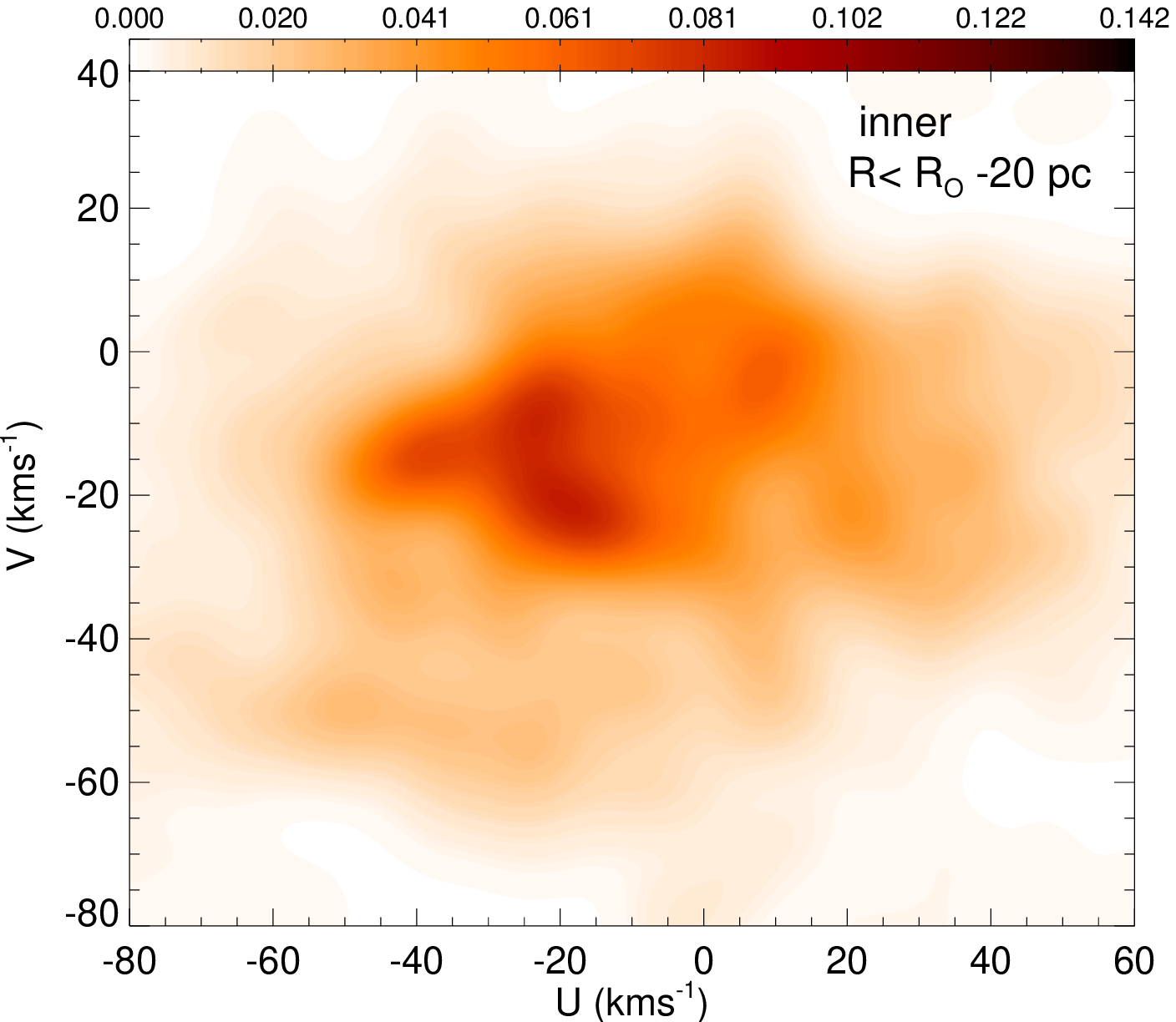}
 \includegraphics[width=0.22\textwidth]{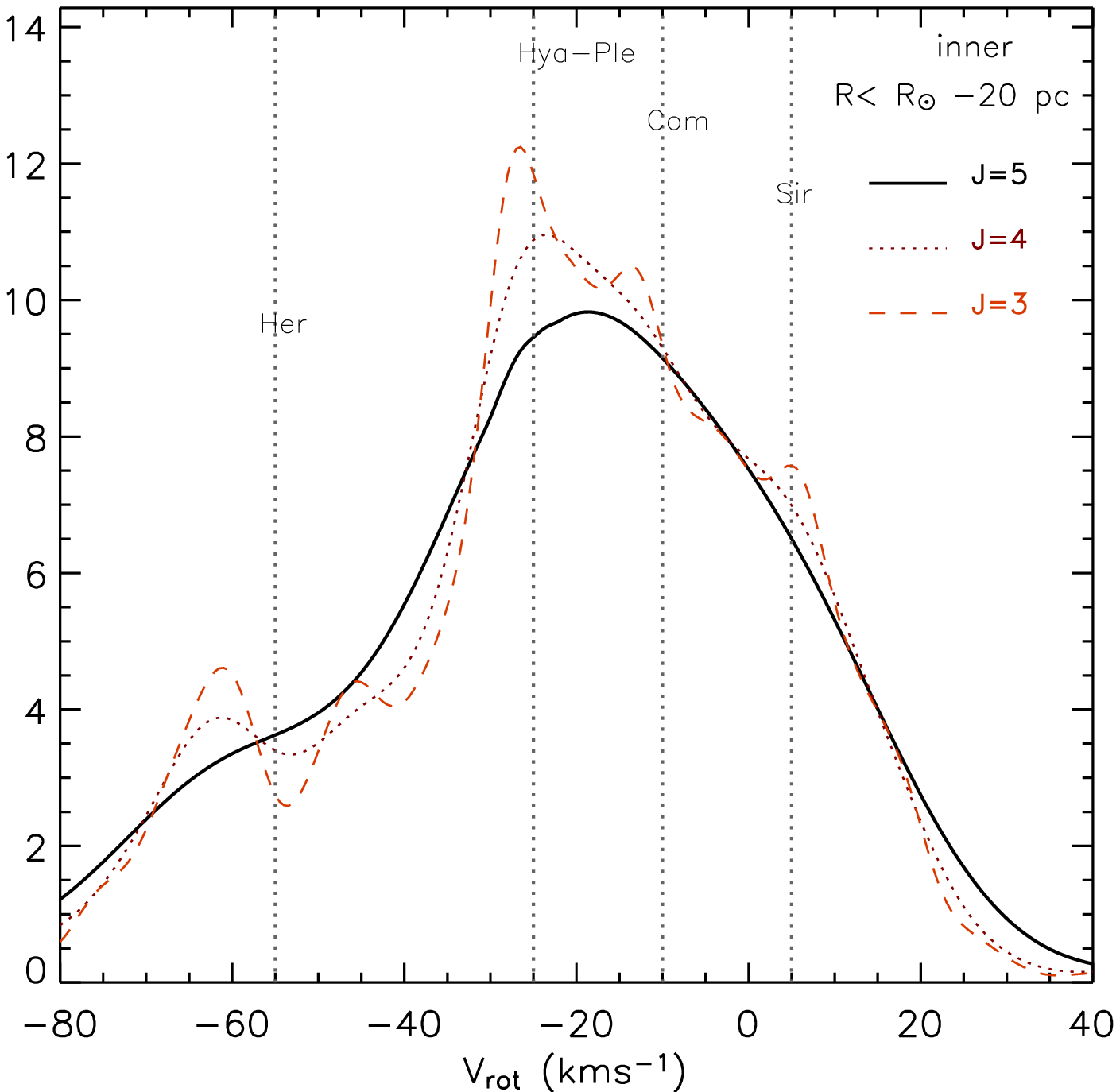}
\includegraphics[width=0.26\textwidth]{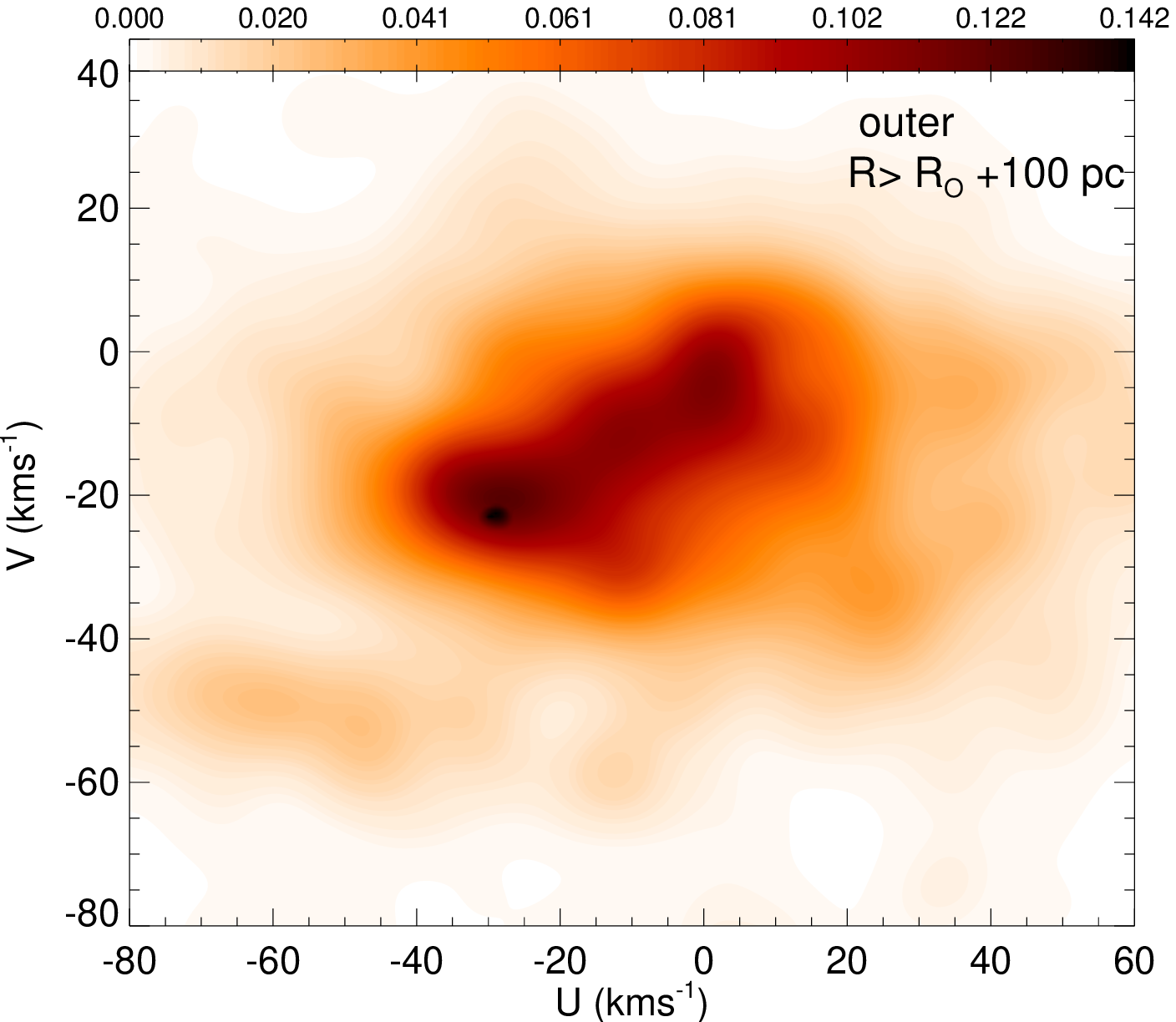}
 \includegraphics[width=0.22\textwidth]{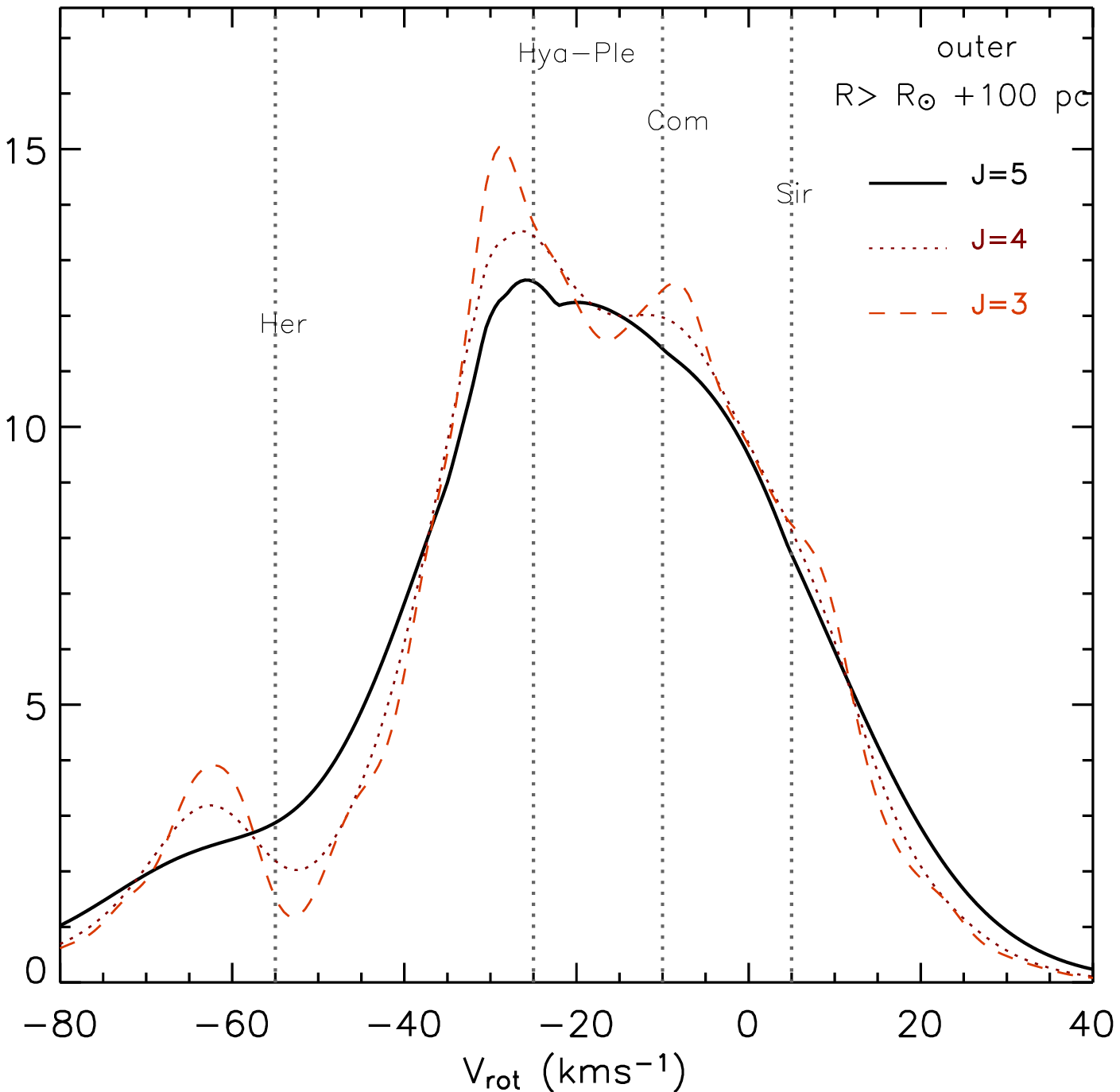}
 \caption{{\bf Left column:} Density field in the $U$--$V$ plane of two subsamples of KM giant stars situated at different galactocentric radii obtained by WD with $J=4$: inner subsample ($R<R_\odot-20\pc$, 1812 stars) and outer subsample ($R\geq R_\odot+100\pc$, 2057 stars). {\bf Right column:} Density field for the $V_{rot}$ component obtained integrating the density in the $U_{rot}$--$V_{rot}$ plane with different values of $J$ for these subsamples.}
    \label{UVradii}
\end{figure}

\begin{figure}
    \centering
 \includegraphics[width=0.26\textwidth]{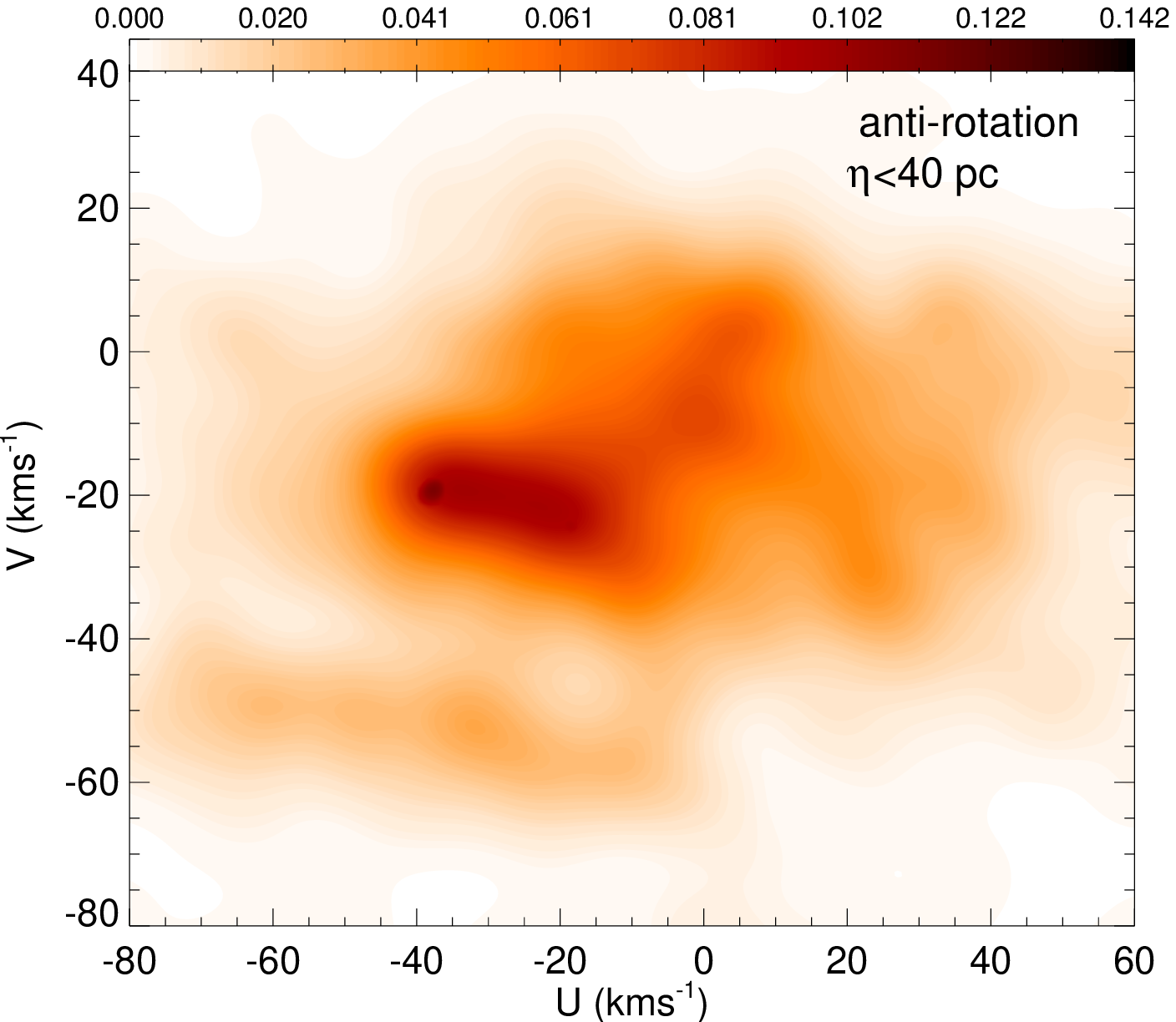}
 \includegraphics[width=0.22\textwidth]{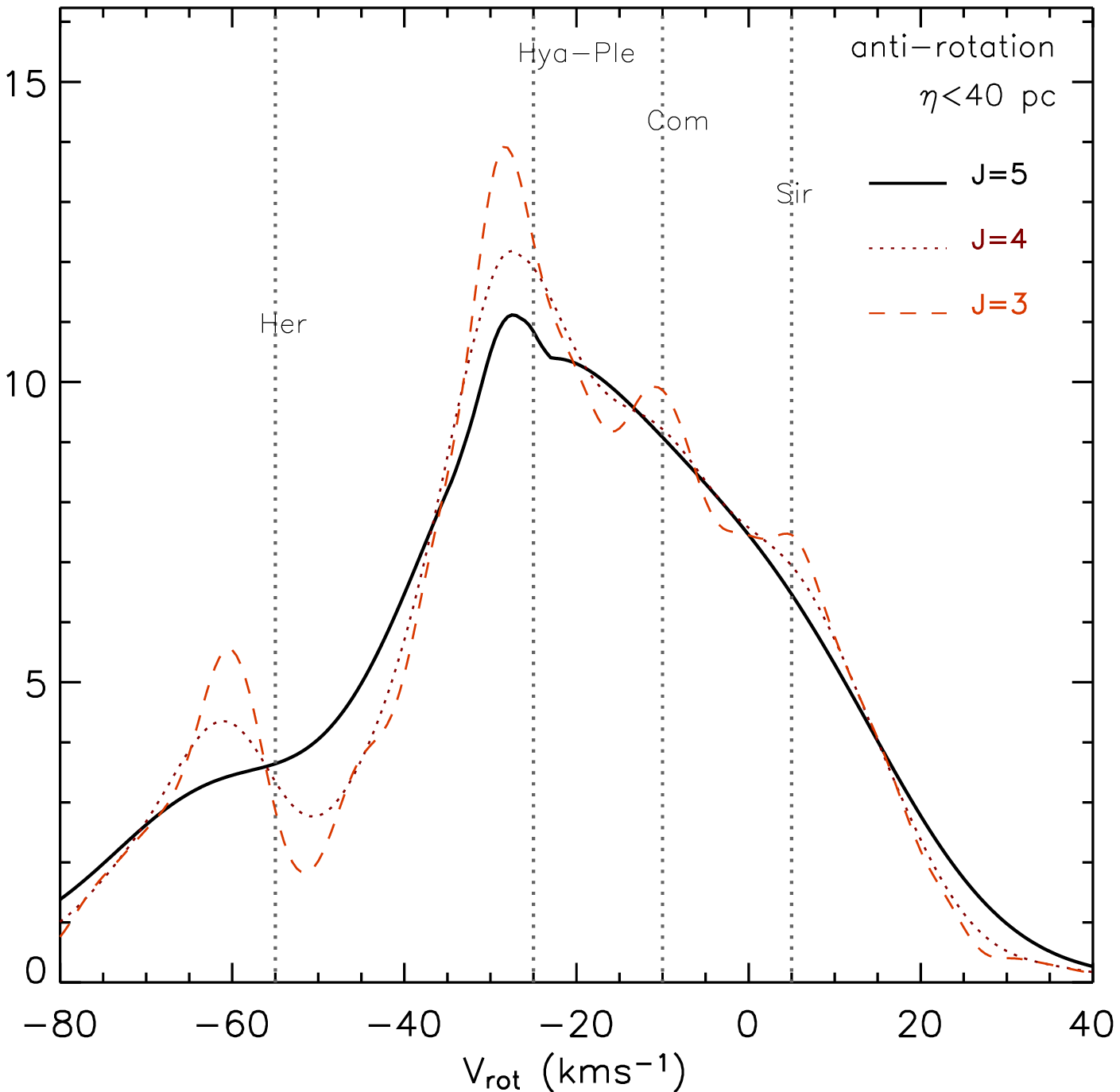}
 \includegraphics[width=0.26\textwidth]{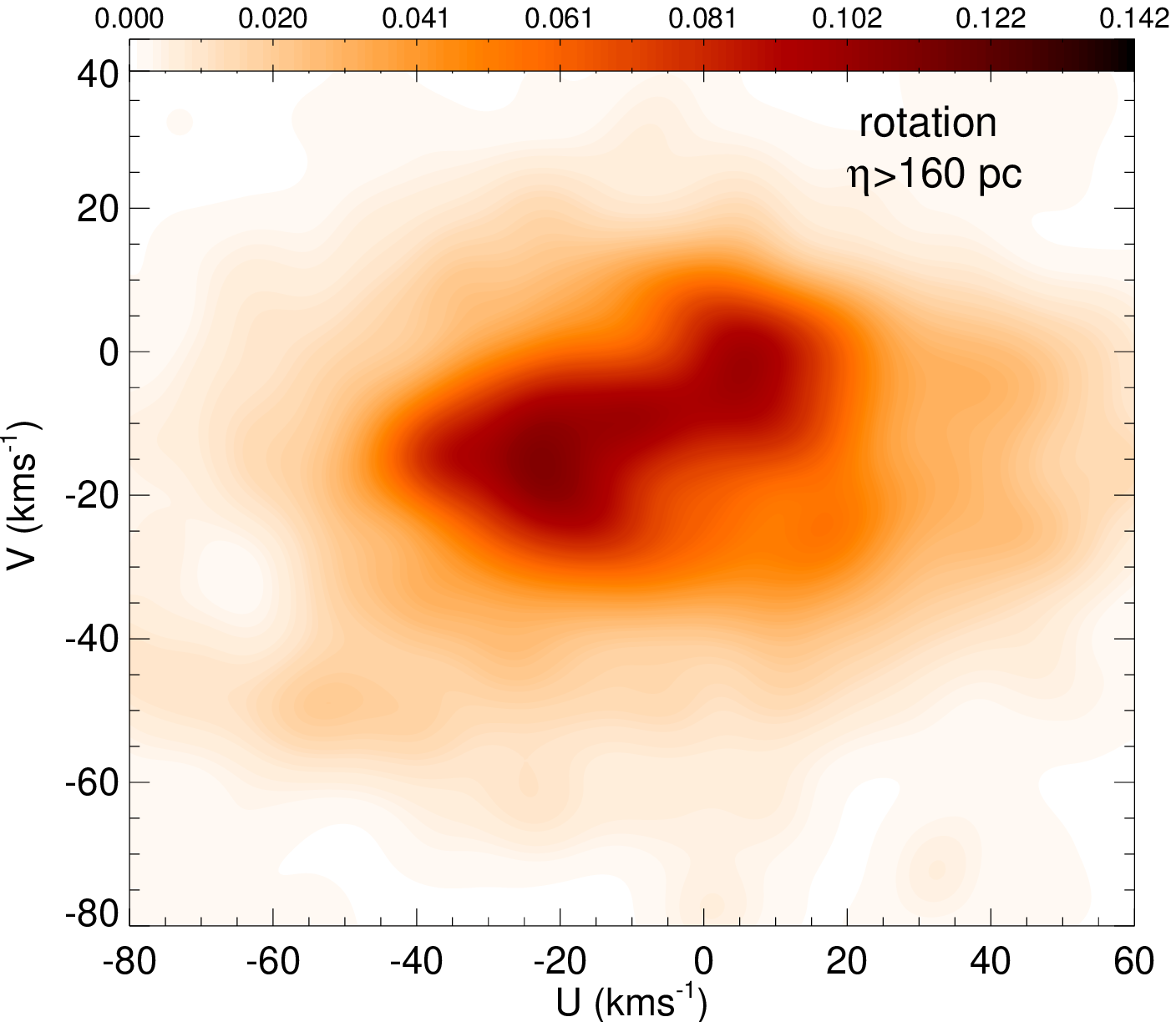}
 \includegraphics[width=0.22\textwidth]{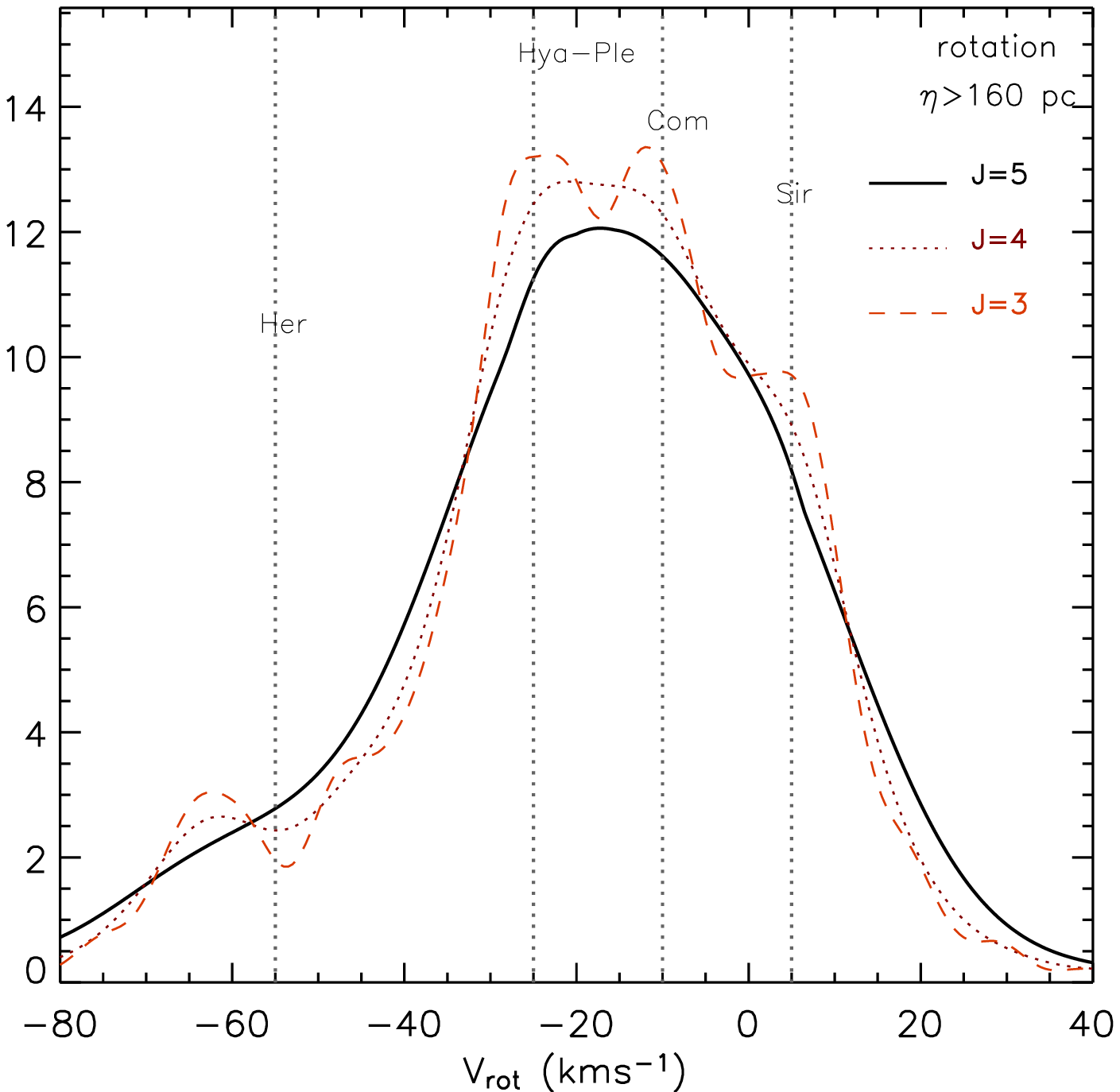}
\caption{{\bf Left column:} Density field in the $U$--$V$ plane of two subsamples of KM giant stars situated at different azimuths obtained by WD with $J=4$: subsample towards anti-rotation ($\eta<40\pc$, 1903 stars) and subsample towards rotation ($\eta>160\pc$, 1975 stars). {\bf Right column:} Density field for the $V_{rot}$ component obtained integrating the density in the $U_{rot}$--$V_{rot}$ plane with different values of $J$ for these subsamples.}
 \label{UVazimuts}
\end{figure}

We want to evaluate whether the subsample of M dwarfs could contribute to solve this issue, despite being aware that the low number of stars in this case prevents us from giving any conclusive statement. The $V_{rot}$ distribution for these stars (Fig. \ref{spectraltypes}) shows two peaks around the Hercules branch at $V_{rot}\sim-50\kms$ and $-65\kms$. As this subsample is composed of stars from two different catalogues (see Sect. \ref{data}), we ascertain that the stars from \citet{reid02} (very nearby stars with $d<25\pc$) show a very broad Hercules branch but centred at $V_{rot}\sim-55\kms$, in agreement with the previous results for nearby stars. However, the stars from \citet{bochanski05} (stars from selected areas mostly at $\delta>0$ and with larger distances up to $\sim 150\pc$) contribute to the split of this branch into two peaks. Again, the importance of the mean distance and spatial distribution of the subsample considered is revealed.

We look for possible variations of the kinematic properties of the branches with Galactic position. At present, the KM giant subsample is the only one with enough stars and space volume coverage to undertake this study. We have selected 4 subsamples at different locations and approximately with the same number of stars (1800-2000 stars). Note that the centre of KM giants spatial distribution onto the Galactic plane is placed 100 $\pc$ away from the Sun in the direction of Galactic rotation and 40 $\pc$ towards the Galactic anti-centre\footnote{Note also that the observational spatial restriction affects the distribution of the subsamples in the direction perpendicular to the Galactic plane.}. Stars in the central region are not considered in order to emphasize the properties of the extremes. First, we divide the sample into two subsamples according to their different galactocentric radii, $R$: $R<R_\odot-20\pc$ (inner subsample) and $R\geq R_\odot+100\pc$ (outer subsample). Secondly, we build two subsamples with different $\eta$ (heliocentric Cartesian coordinate towards the direction of the Galactic rotation): with $\eta<40\pc$ (subsample towards anti-rotation with respect to the centre of the sample) and $\eta>160\pc$ (subsample towards rotation).

 The results are shown in Figs. \ref{UVradii} and \ref{UVazimuts}. The same color scale is used in all cases for a clearer comparison. From these figures we conclude that the four branches are present in all regions. Second, the Hyades-Pleiades branch is the dominant structure, except for the region towards Galactic rotation where Coma Berenices has the same density. Furthermore, a significant change of contrast among substructures inside the branches is confirmed and the density maximum along the branches (along the $U_{rot}$ component) varies for each region. Note for instance the substructures at $(U,V)\sim(-27,-22)\kms$ and $(U,V)\sim (-20,-10)\kms$ or the Pleiades moving group. Also the shape of the Hercules branch changes between regions and it is more significant for the region through anti-rotation. All these considerations suggest a real effect of Galactic position on the shape of kinematic structures, independently of the possible bias in the LM distances. 

Resuming the issue of the discrepancies in the Hercules branch position, we observe that this branch appears at $V_{rot}\sim-62\kms$ for all these four different galactocentric directions of the KM giants subsample, in strong contrast with the peak at $V_{rot}\sim-55\kms$ for the central nearby stars of this same subsample. This makes the method for LM distance derivation suggestive of a possible bias entangling kinematics and distances in a complex way. Although the Bayesian parametric LM approach was developed to derive unbiased distances, the a priori adoption of a Schwarzchild ellipsoid for the velocity distribution functions can directly affect the distance estimate of a given kinematic group or branch. Further work along these lines requires the revision of the LM distances and the application of the WD method to larger and spatially more extended surveys (RAVE, Gaia).


\section{Age--kinematics characterization}\label{age}

Two different approaches are adopted in this section. First, the whole sample is divided by age range into statistically significant subsamples to trace the kinematic structures in the $U$--$V$ plane through age.
Secondly, only those stars with the most precise ages --relative errors of less than 30\%-- are used to study structures and periodicities in $U$--$V$--$age$ space and other connections between kinematic structures and the evolutionary state of their members in the context of the branches proposed in Sect. \ref{branches}. Note that in both analyses the sample of stars used only includes the FGK and OBA dwarfs samples, since individual ages are not available for the other subsamples (see Table \ref{tab.data})\footnote{In Sect. \ref{data} we mention that the H$\alpha$ equivalent width could be used as information on age for the M dwarfs. We have checked, however, that the number of stars per sample when active and non-active stars are analysed separately is too low to derive significant results from the distributions: too smoothed distributions are obtained by WD with $J_{plateau}$. This study has been postponed until the use of recent new data, i.e. \citet{bochanski07}.}.

\subsection{Age dependence in the $U$--$V$ plane}\label{age1}

\begin{figure}
    \centering
\includegraphics[width=0.24\textwidth]{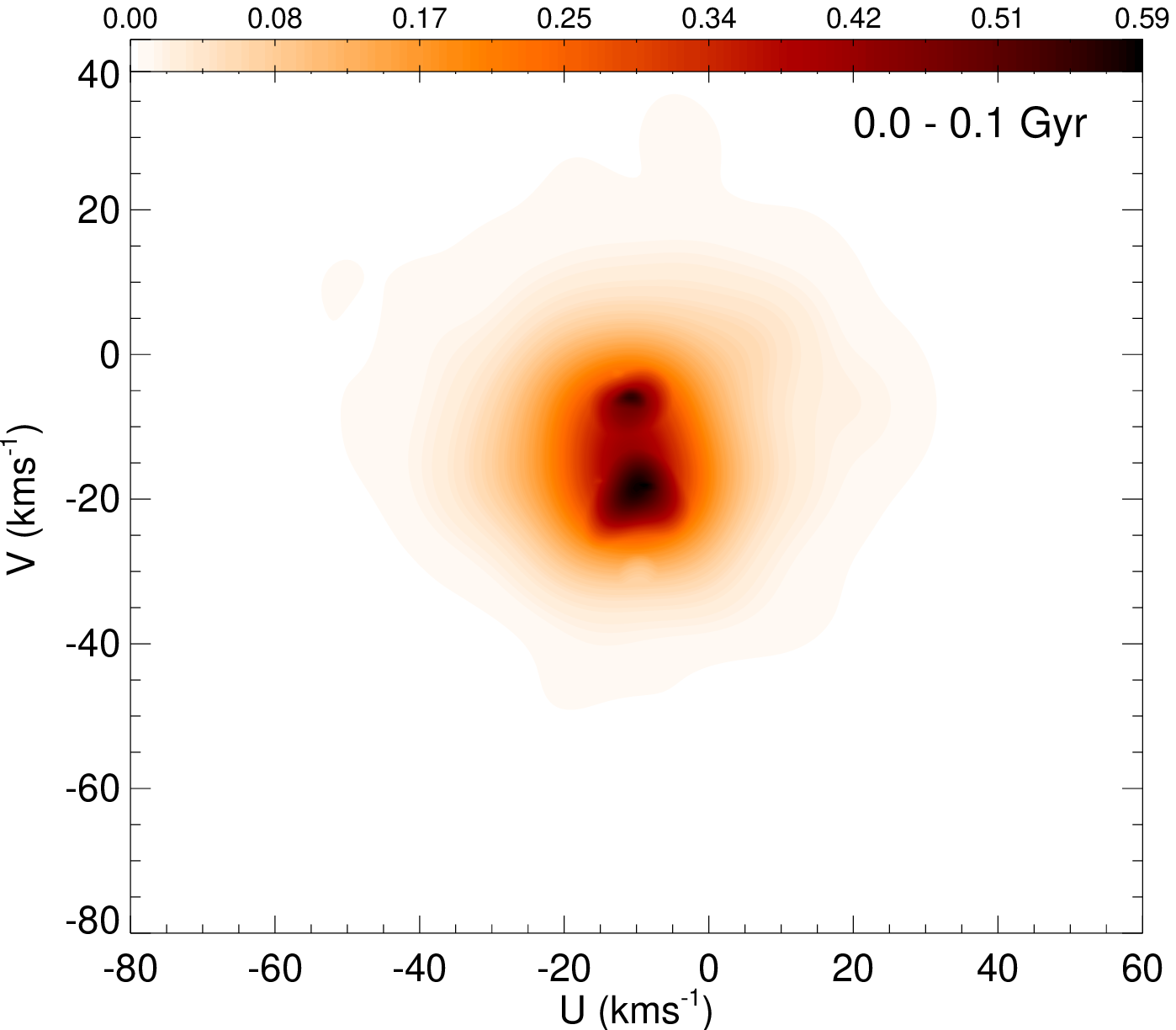}
\includegraphics[width=0.21\textwidth]{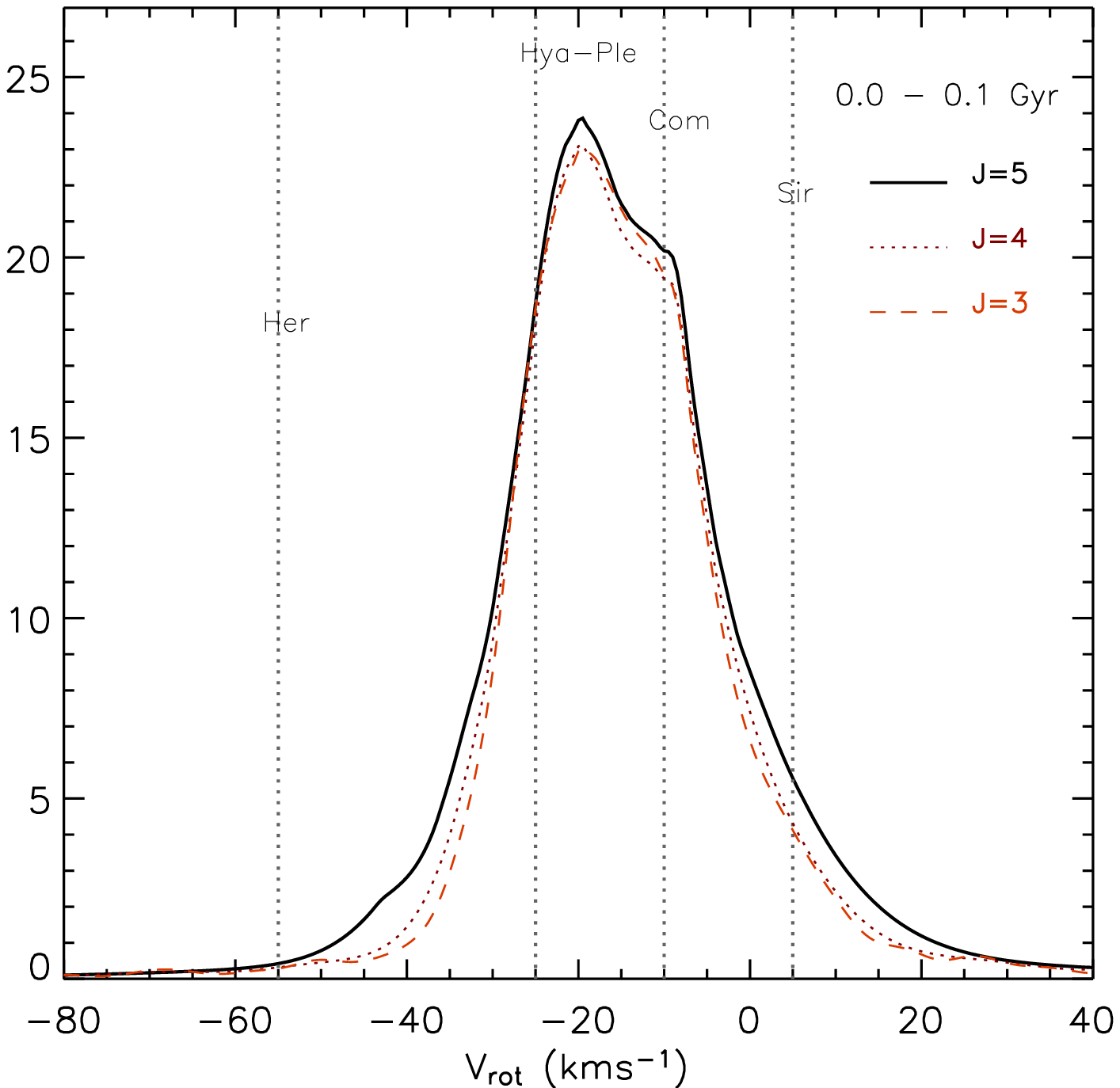}

\includegraphics[width=0.24\textwidth]{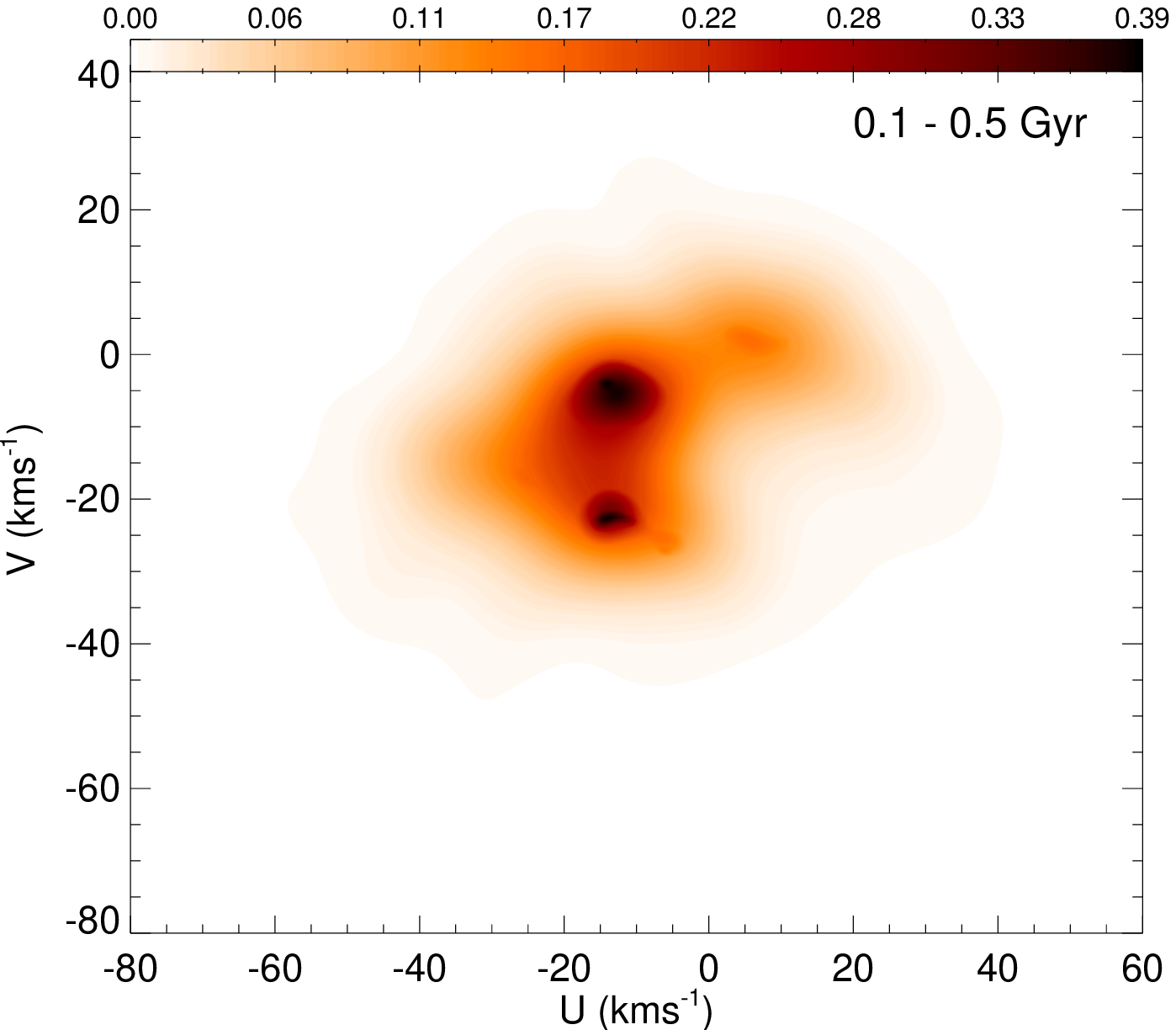}
\includegraphics[width=0.21\textwidth]{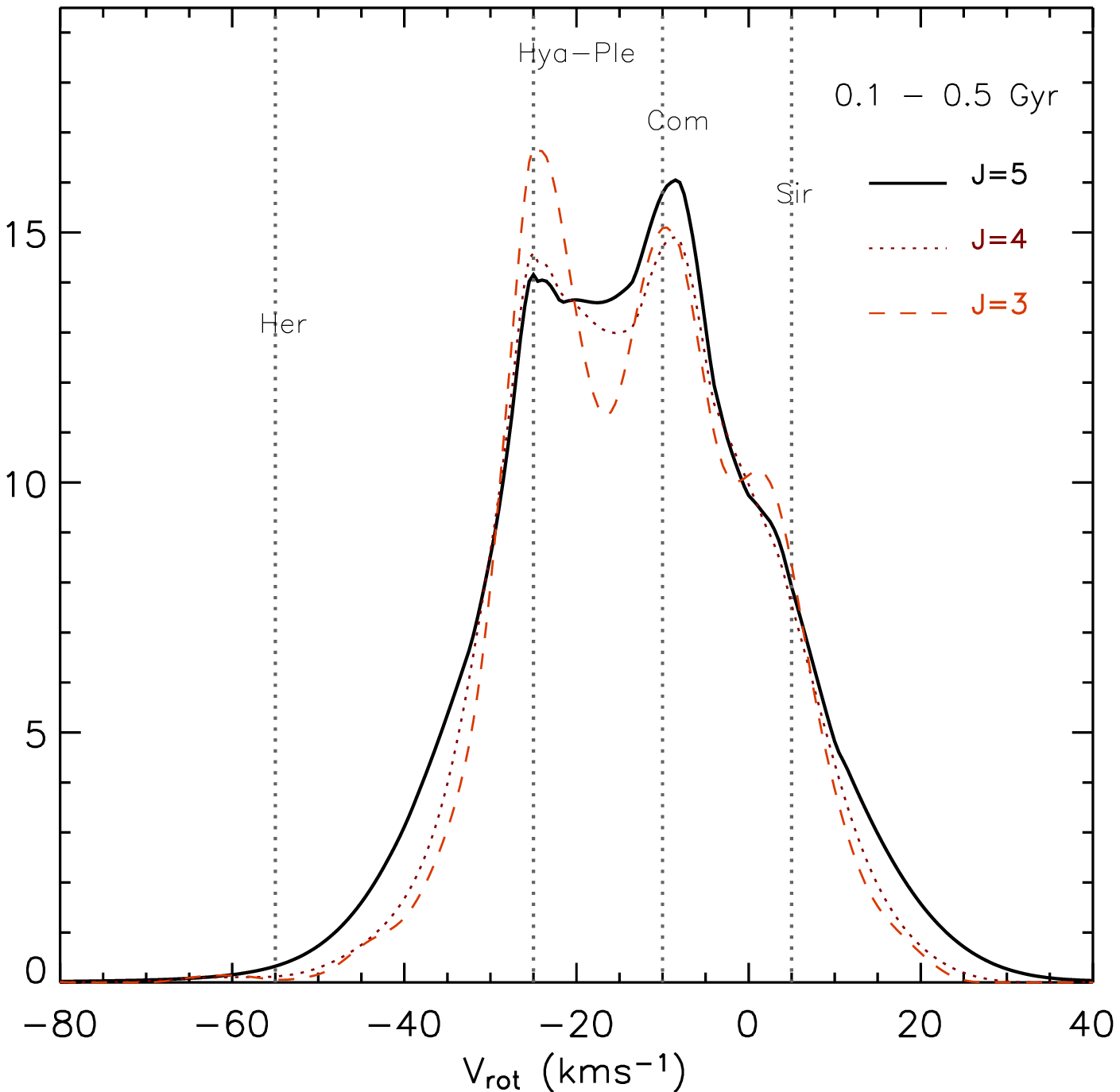}

\includegraphics[width=0.24\textwidth]{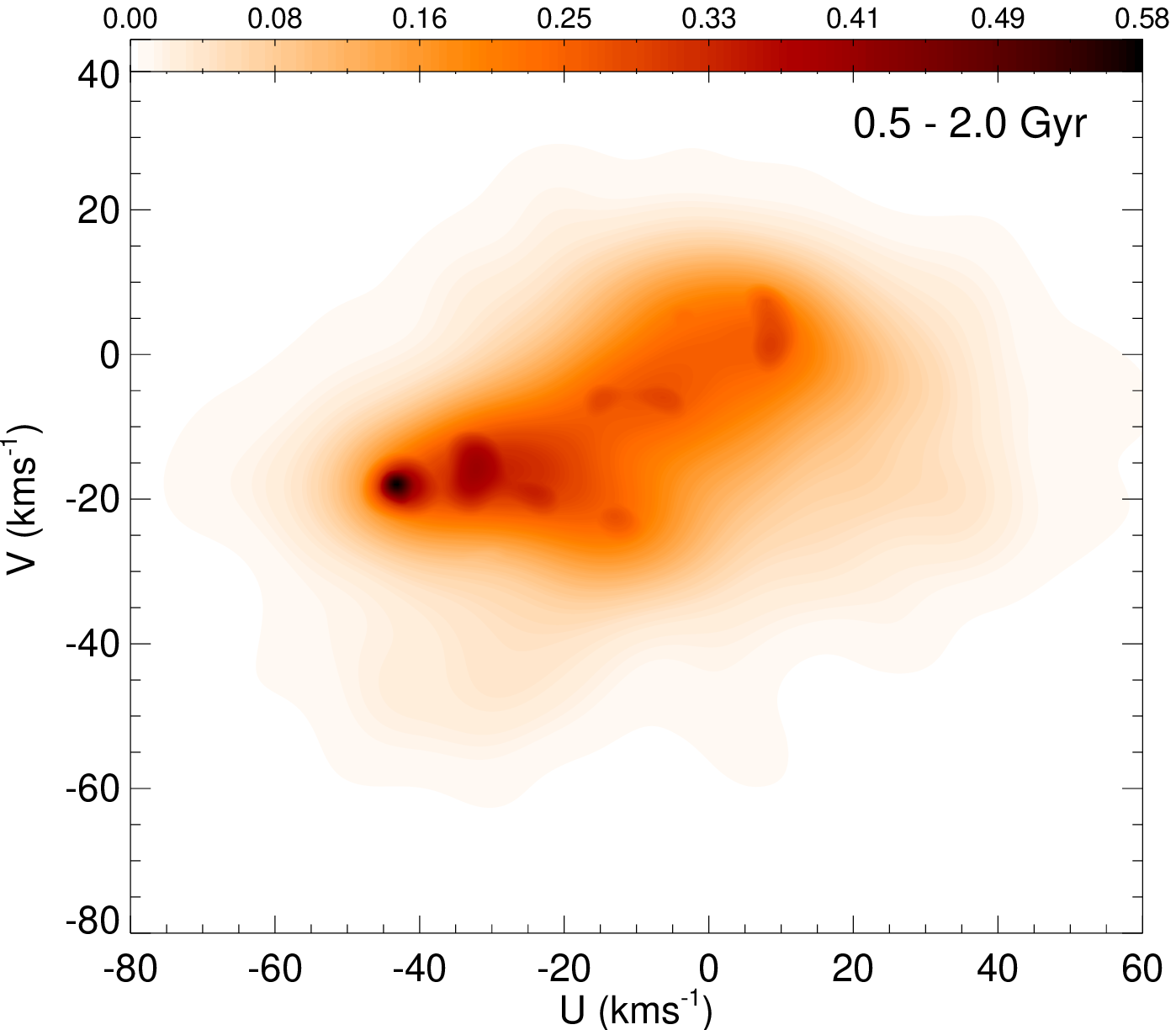}
\includegraphics[width=0.21\textwidth]{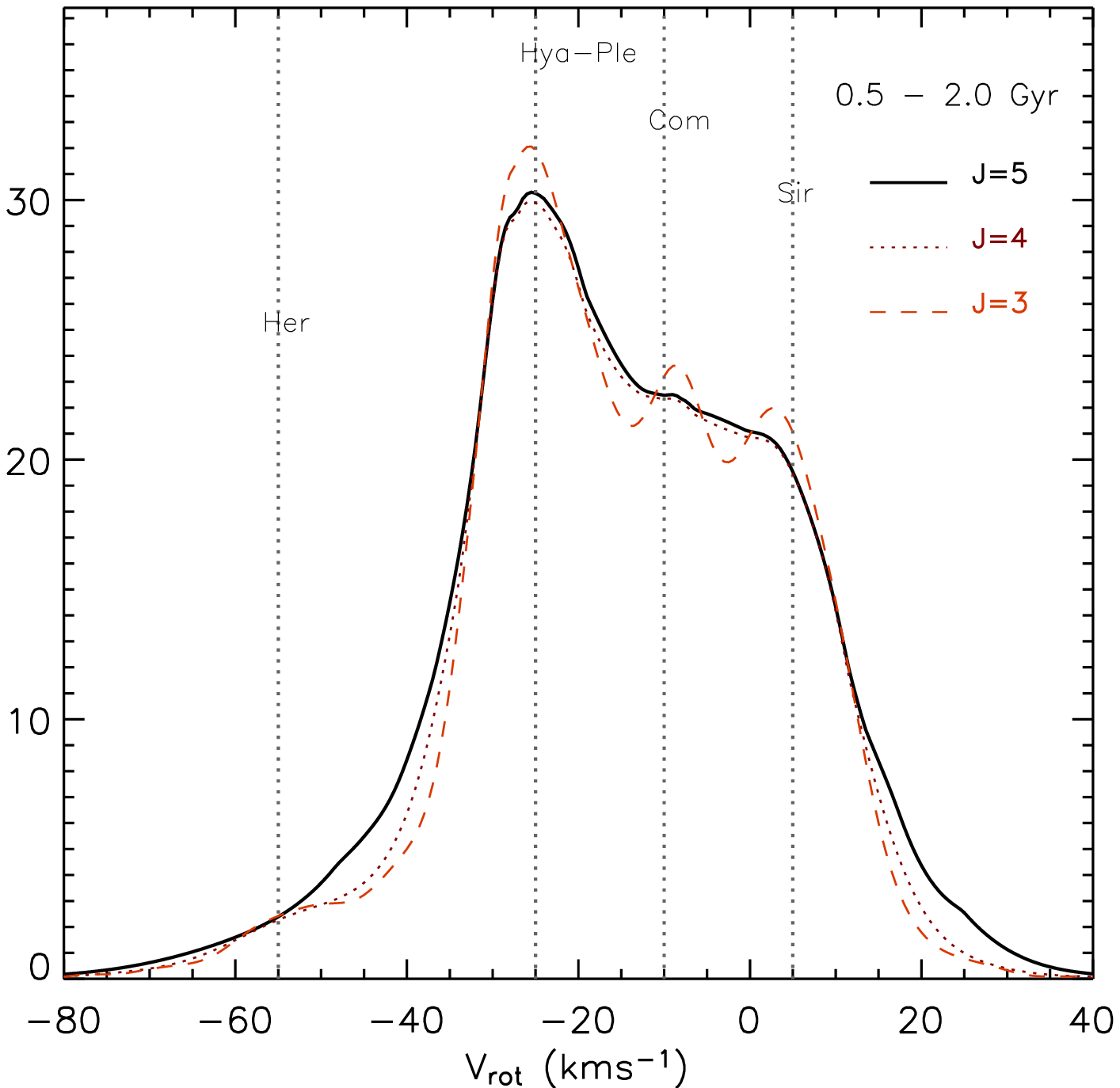}

\includegraphics[width=0.24\textwidth]{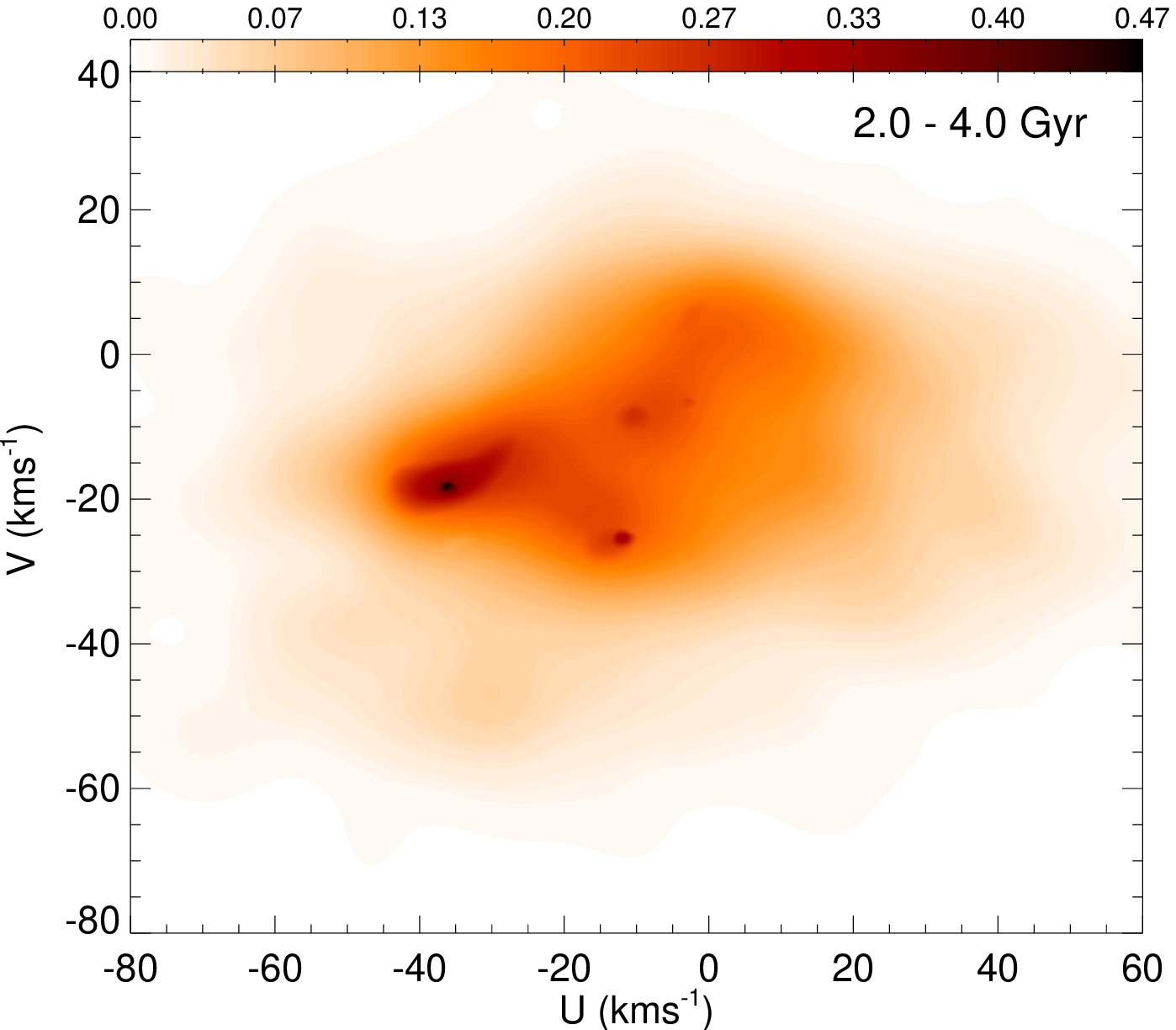}
\includegraphics[width=0.21\textwidth]{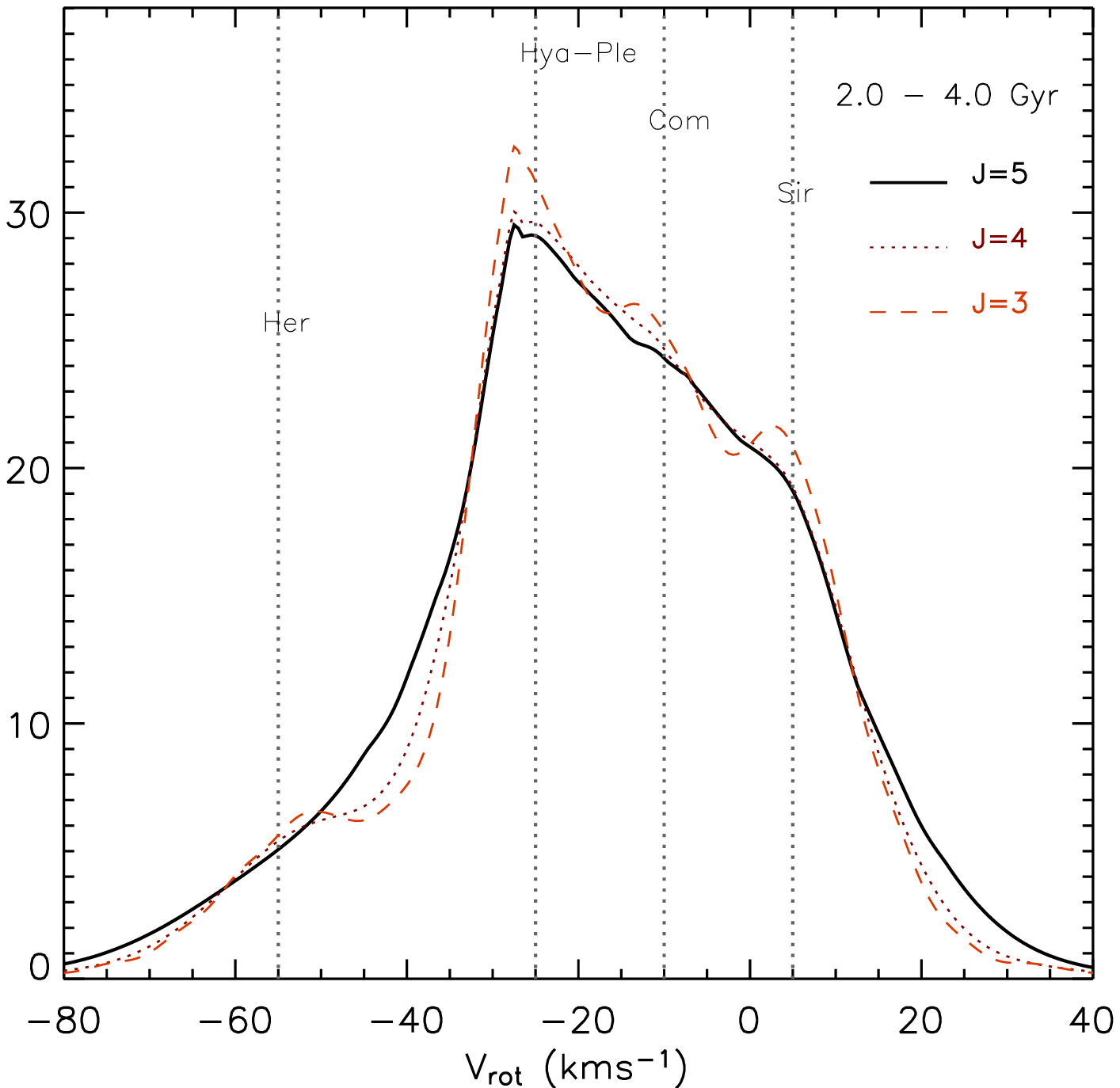}

\includegraphics[width=0.24\textwidth]{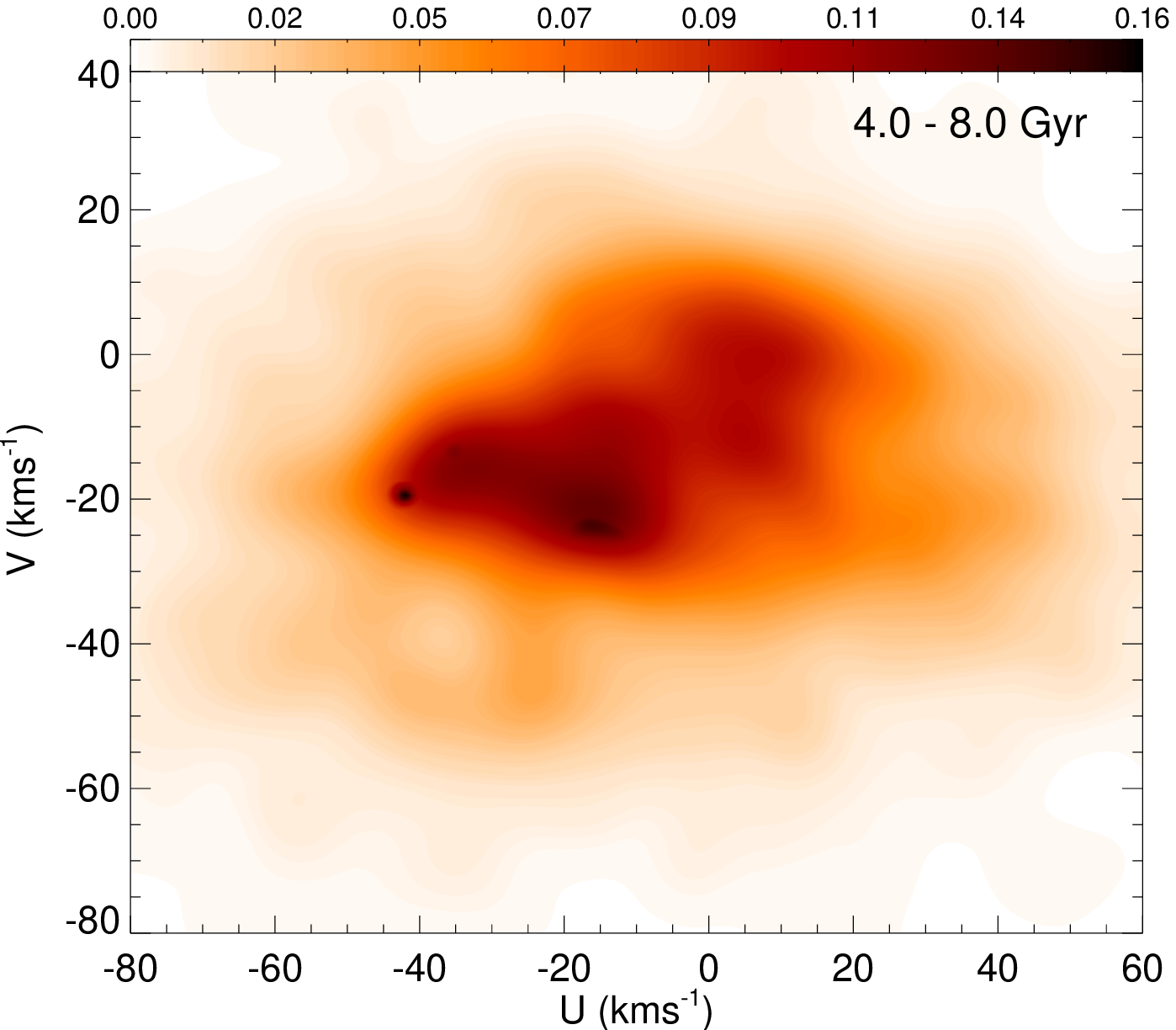}
\includegraphics[width=0.21\textwidth]{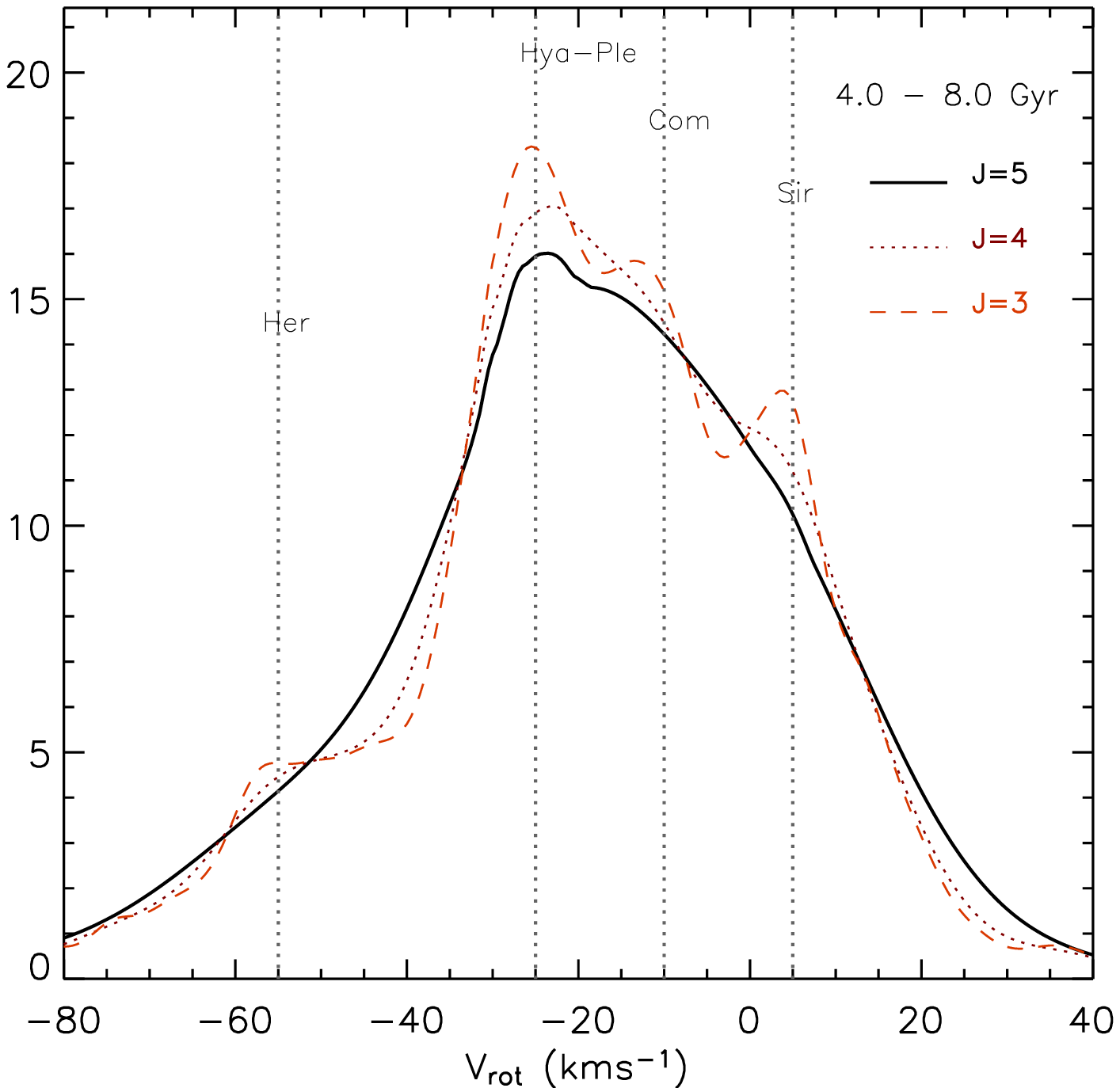}

\includegraphics[width=0.24\textwidth]{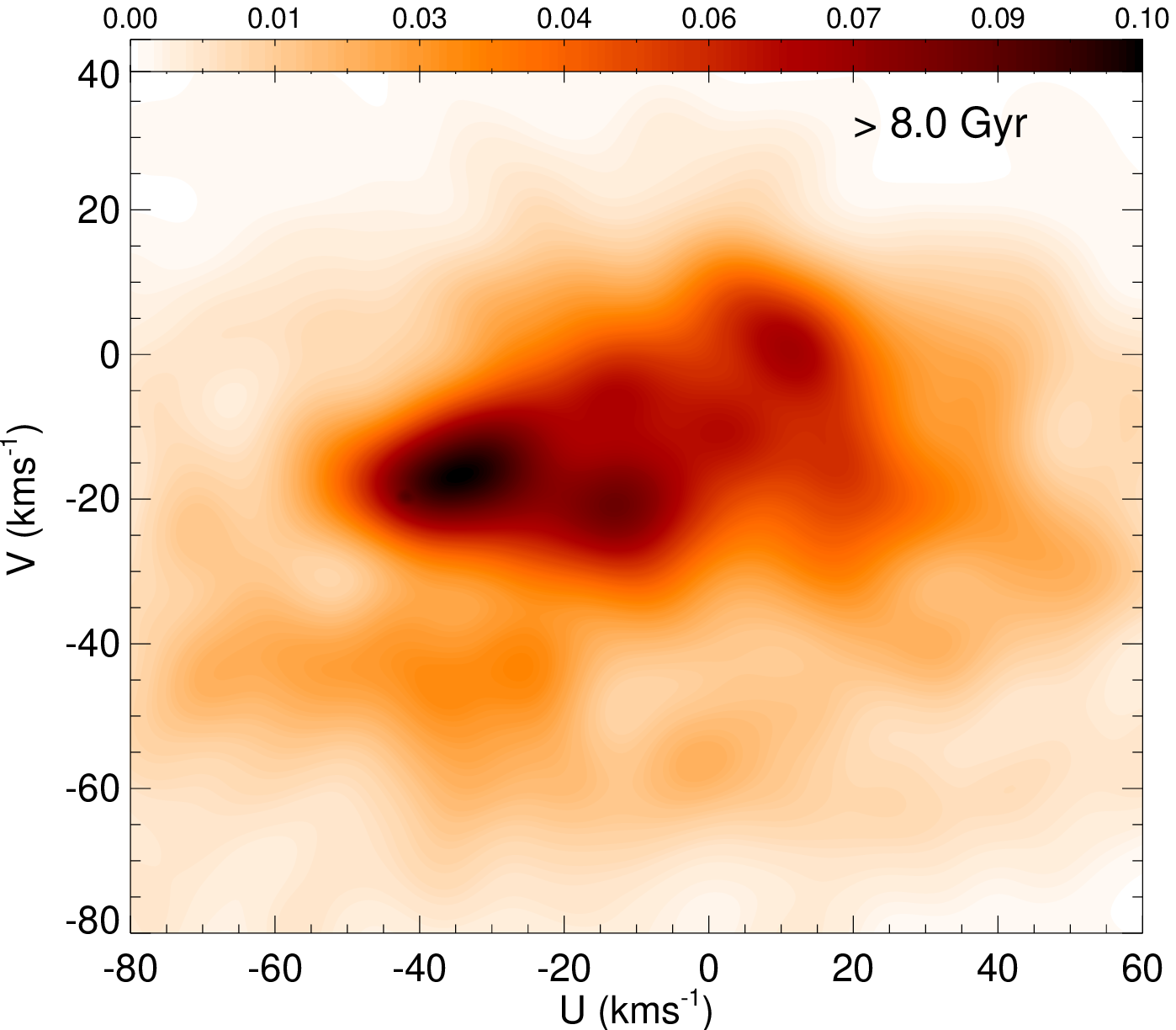}
\includegraphics[width=0.21\textwidth]{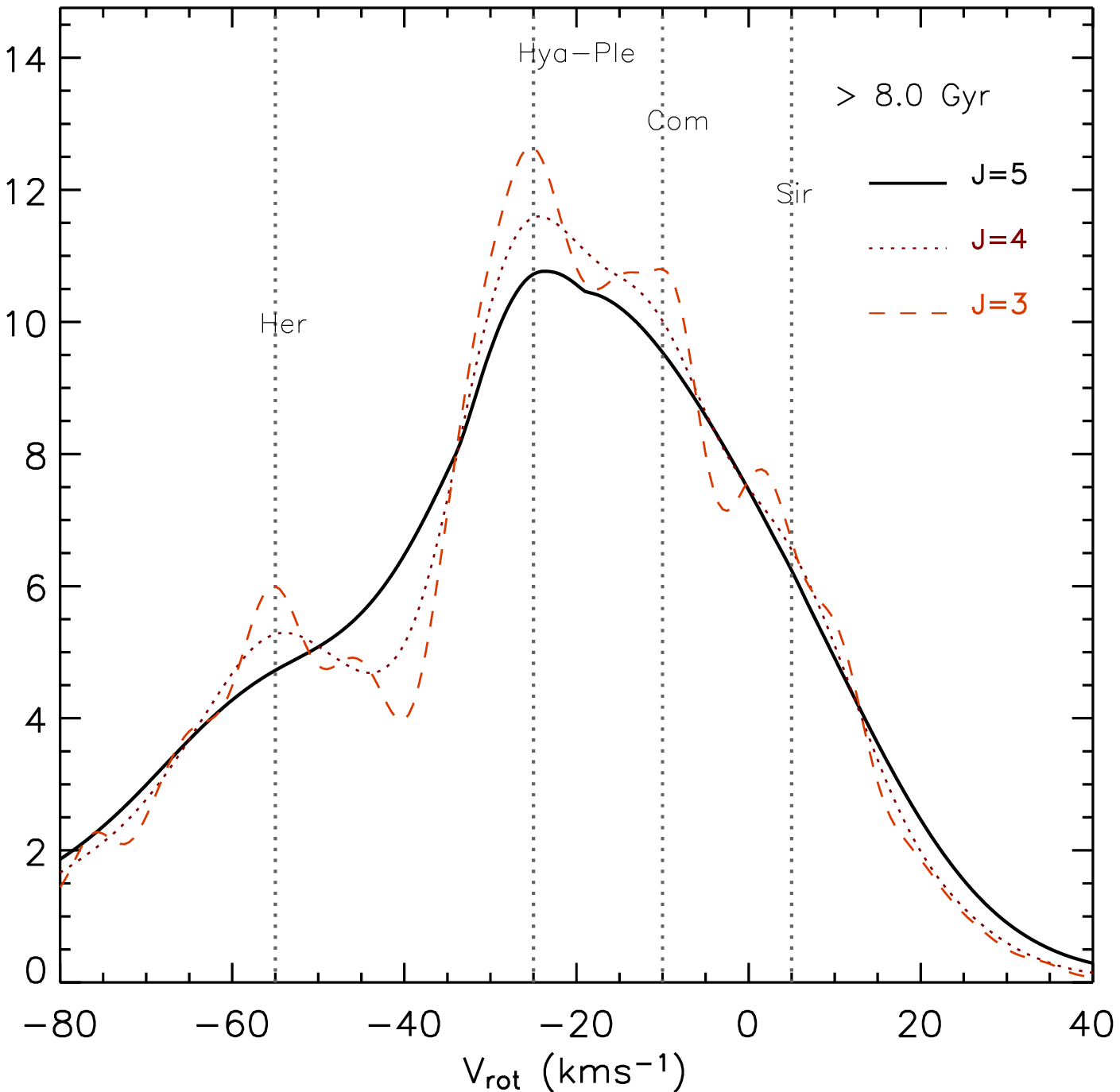}
\caption{{\bf Left column:} Density field in the $U$--$V$ plane obtained by WD with $J=4$ for subsamples of different ages (in $\Gyr$): 0-0.1 (1792 stars), 0.1-0.5 (1501), 0.5-2.0 (3368), 2.0-4.0 (3917), 4.0-8.0 (2561) and $>$8.0 (2053).  {\bf Right column:} Density field for the $V_{rot}$ component obtained integrating the density in the $U_{rot}$--$V_{rot}$ plane and with different values of $J$ of the WD: 3, 4 and 5.}
\label{UV_age}
\end{figure}

\begin{figure*}
    \centering 
\includegraphics[width=0.45\textwidth]{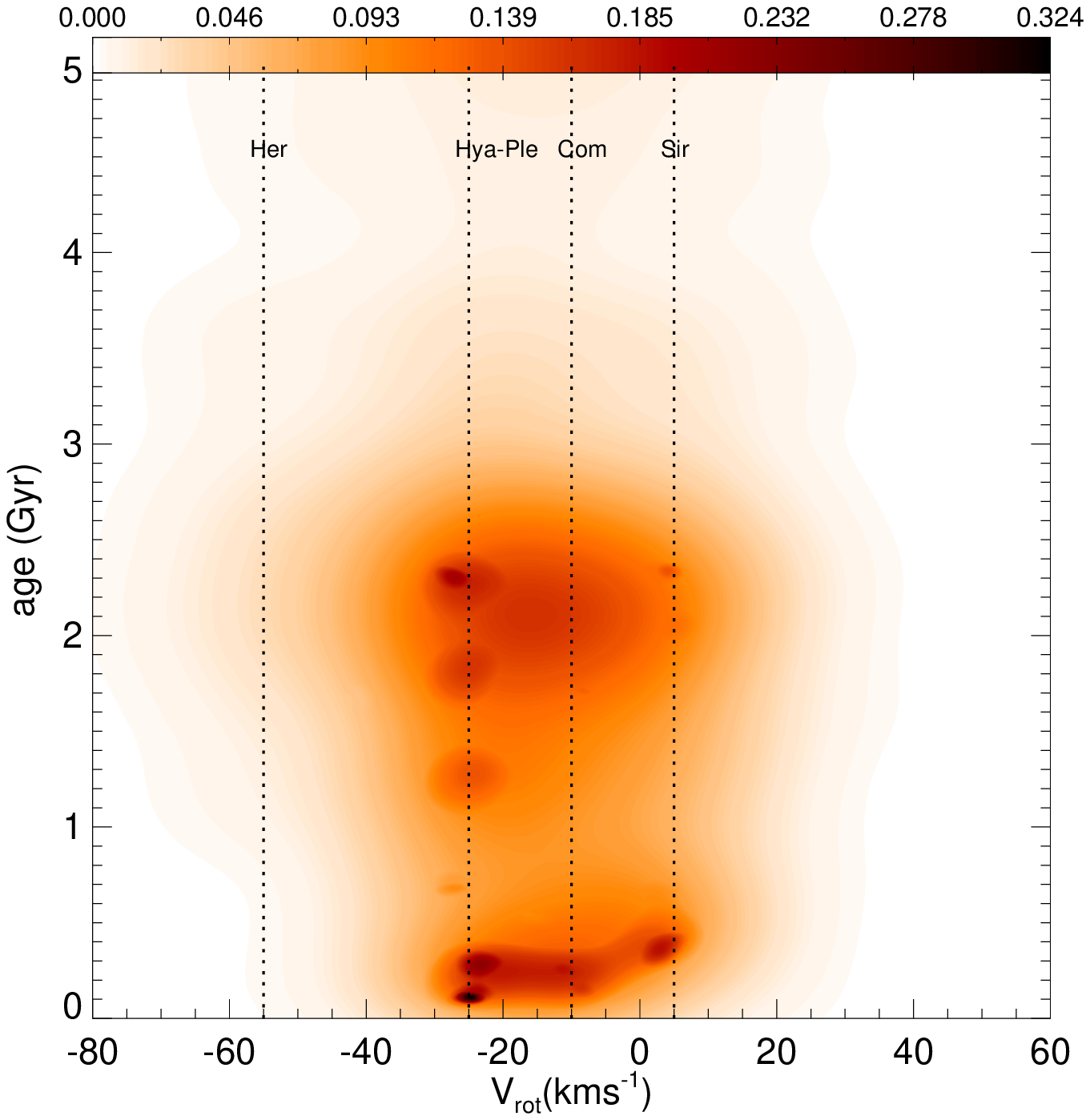}
\includegraphics[width=0.4\textwidth]{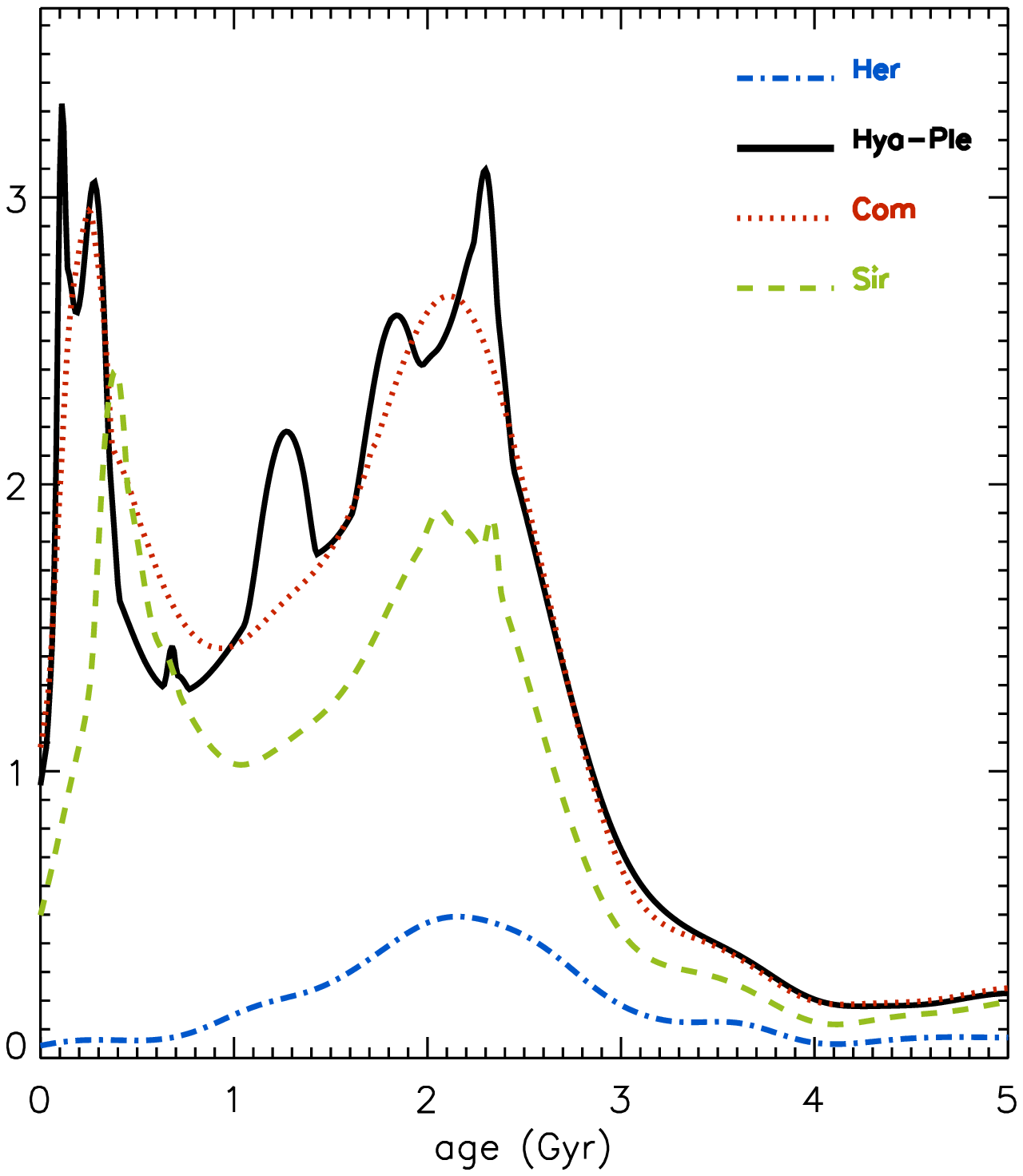}
    \caption{{\bf Left:} Density field in the $V_{rot}$--$age$ plane of the stars with $\epsilon_{age}\leq 30\%$  (7016 stars) obtained by WD. {\bf Right:} Integrated density of stars in each branch worked out using the WD method as a function of age for stars with $\epsilon_{age}\leq 30\%$ (223, 1195, 1078 and 806 stars in the Hercules, Hyades-Pleiades, Coma Berenices and Sirius branches respectively).} 
    \label{V_rot_age}
\end{figure*}

Figure \ref{UV_age} shows the distribution in the $U$--$V$ plane and the distribution for the $V_{rot}$ component obtained by WD for subsamples with different age ranges. The age bins have been chosen in order to emphasize the changes in these distributions. Important observations derived from Fig. \ref{UV_age} are:

\begin{itemize}

\item For the youngest stars, only the Pleiades and Coma Berenices kinematic groups appear and the branches are not identified in the $U$--$V$ plane. For ages between $0.1$ and $0.5\Gyr$, the Hyades and Sirius structures have begun to appear. The Hyades structure is the most prominent in all the subsamples with age $>0.5\Gyr$.

\item Certain structures appear and disappear (or at least their contrasts vary considerably in relation to other structures) when subsamples with different age ranges are considered. See, for instance, the Pleiades structure at $(U,V)\sim(-12,-22)\kms$. From top to bottom, we see how this structure is clearly evident for very young stars but is less significant in the $0.5$-$2.0$ $\Gyr$ range, only to reemerge at 4-8 $\Gyr$\footnote{\citet{asiain99b}, working only with the sample of B and A main sequence stars, found a significant variation of the Pleiades moving group with age. However, as ages in their sample are $<0.5\Gyr$, their interpretations cannot be applied here.}.

\item The structure of branches is well traced up to the oldest samples. The minimum age of the stars in each branch is discussed in Sect. \ref{age2}.

\item The separation between branches does not depend on age. The value of 15$\kms$ is found to be appropriate except for the youngest stars ($<$100$\Myr$) where the Pleiades and Coma Berenices moving groups are closer in $V_{rot}$.

\item The drop of density at the left of the Hyades-Pleiades branch considered in Sect. \ref{branches} is noticed in all age ranges. In addition, we observe that the relative density between this branch and the Hercules branch clearly decreases with age.

\end{itemize}

\subsection{Structures and periodicities in the $V_{rot}$--$age$ plane}\label{age2}

Figure \ref{V_rot_age} (left) shows the distribution in the $V_{rot}$--$age$ plane for the stars with well-defined ages ($\epsilon_{age}\leq 30\%$, 7016 stars). The age distribution of each branch (Fig. \ref{V_rot_age}, right) is computed from the whole distribution, selecting the region of the branches as detailed in Sect. \ref{branches} (centres at $-55$, $-25$, $-10$ and $5$ $\kms$ for the Hercules, Hyades-Pleiades, Coma Berenices and Sirius branches respectively with a width of $\pm 4\kms$). Important observations derived from this distributions are:

\begin{itemize}

\item An extended age distribution is confirmed for all four branches.

\item While the Hyades-Pleiades and Coma Berenices branches have an important fraction of very young stars, Sirius has its first main peak for slightly older stars at $\sim400\Myr$. Consequently, the velocity distribution becomes structured by the tree main branches for stars of $\sim 400\Myr$.

\item \citet{dehnen00} assigns an age of $>8 \Gyr$ to the Hercules group and \citet{bobylev07} claim that it does not show any branch-like shape unless very old ($>8 \Gyr$) subsamples are studied. Here the kinematic region of the Hercules branch is populated with stars of ages of more than $1 \Gyr$ and its extended branch-like shape is seen in all subsamples with ages $>2\Gyr$ (Fig. \ref{UV_age}) where it reaches to a quite extended longitude.

\item A clumpy distribution is observed inside the Hyades-Pleiades branch, with a periodicity in age of about 500-600 $\Myr$, maintained at least in the $0$-$2.5 \Gyr$ range. However, this periodicity is marginally significant\footnote{Notice that the periodicity is statistically significant in the sense of the discussion in Sect.\ref{method} ($J=J_{plateau}$).} as the absolute error in age of some stars exceeds the period obtained. We point out that this periodicity should be contrasted in the future with the star formation rate obtained in \citet{hernandez00}, where an oscillatory component of period $\sim 500 \Myr$ is found. For the other branches only an outline of the shape of the whole age distribution is observed.

\end{itemize}

\section{An attempt to evaluate $[Fe/H]$ dependence}\label{feh}

The study of the $[Fe/H]$ metallicity distribution of the stars along the branches and its relation to kinematics and age is restricted to the FGK dwarfs, which is the only sample for which metallicity data is available (see Table \ref{tab.data}). 
In Table \ref{met_grups} we present the mean metallicity and dispersion values for each branch with stars selected as discussed in Sect. \ref{branches}. The values obtained by \citet{helmi06}, using the same sample but a different definition of the position of the kinematic structures, are included in the table for comparison (Coma Berenices was not included in Helmi's analysis). These authors found a common metallicity dispersion of $0.2\dex$ for all three superclusters. However, we obtain slightly lower mean metallicities for all the branches and a higher metallicity dispersion for the Hercules branch. \citet{haywood06} pointed out that systematic biases in the data from \citet{nordstrom04} provoke an artificial increase in the $[Fe/H]$ dispersion at least for ages $<3\Gyr$. As the four branches present a wide range of ages (Sect. \ref{age}), it will be assumed that this specific bias influences each of the branches similarly.

\begin{figure}
    \centering
    	\includegraphics[width=0.45\textwidth]{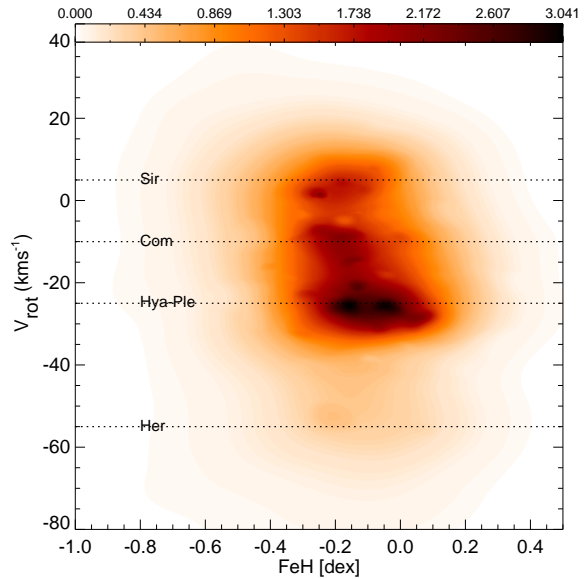}
    \caption{Density field in the $V_{rot}$--$[Fe/H]$ plane for the sample with available $[Fe/H]$ (13109 stars) obtained by WD.}
    \label{Vfeh}
\end{figure}

More importantly, in Fig. \ref{Vfeh} we present the distribution in the $[Fe/H]$--$V_{rot}$ plane. Both the WD method and the rotation of the velocity components by $16\deg$ allow us to identify the four conspicuous branches and their rather wide range of metallicity. A general correlation can be seen: a more negative $V_{rot}$ implies a higher mean metallicity\footnote{Notice that this correlation was not as clearly observed in \citet{nordstrom04} (see their Fig. 32), despite using the same sample, which demonstrates the capabilities of the WD method and the rotation.}. Notice that the correlation is complex due to either the bimodality of the Hyades-Pleiades branch or to the fact that the Hercules branch does not follow the overall pattern of the three main branches.

\begin{figure}
    \centering 
 \includegraphics[width=0.35\textwidth]{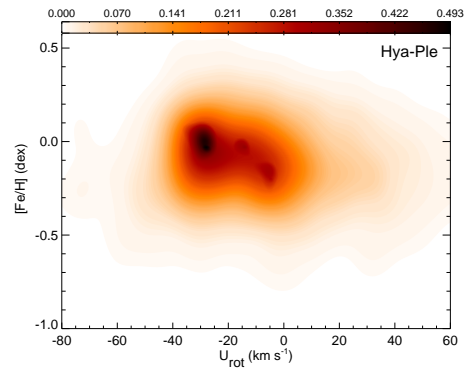}
\caption{Density field in the $U_{rot}$--$[Fe/H]$ plane obtained by WD for the Hyades-Pleiades (2271 stars).}
\label{Ufeh_hyaple}
\end{figure}

 The bimodality for the Hyades-Pleiades branch shows the peaks at $[Fe/H]\sim-0.05$ and $[Fe/H]\sim-0.15$. The metallicity distribution for the stars along the position that this branch traces in the $U$--$V$ plane is studied in detail in Fig. \ref{Ufeh_hyaple}. We observe that the $[Fe/H]$ distribution evolves from the Hyades ($U_{rot}\sim-30\kms$) with the peak at $[Fe/H]\sim-0.05$ --in agreement with the results of \citet{famaey07}-- to the Pleiades, which placed at $U_{rot}\sim-5\kms$ shows a peak at $[Fe/H]\sim-0.15$.
 
\begin{table}
\begin{center}
\caption{Mean metallicities and dispersions for the branches considered in the present work and for the superclusters according to (1) \citet{helmi06}.}
\label{met_grups}
\begin{tabular}{lcc} 
\hline \hline
                & Branches (this work)                & Superclusters (1)\\
                &$\overline {[Fe/H]}\ \ $ $\sigma_{[Fe/H]}$&$\overline {[Fe/H]}\ \ $ $\sigma_{[Fe/H]}$  \\
                &($\dex$)$\ \ \ $ ($\dex$)&($\dex$)$\ \ \ $ ($\dex$) \\ \hline
Hercules        & $-0.15\ \     $  0.27   & $-0.13\ \      $0.2\\ 
Hyades-Pleiades & $-0.11\ \     $  0.20   & $-0.08\ \      $0.2\\
Coma Berenices  & $-0.16\ \     $  0.22   &  ...     \\
Sirius          & $-0.21\ \     $  0.21   & $-0.18\ \      $0.2\\
\hline
\end{tabular}
\end{center}
\end{table}

\begin{figure}
    \centering 
\includegraphics[width=0.35\textwidth]{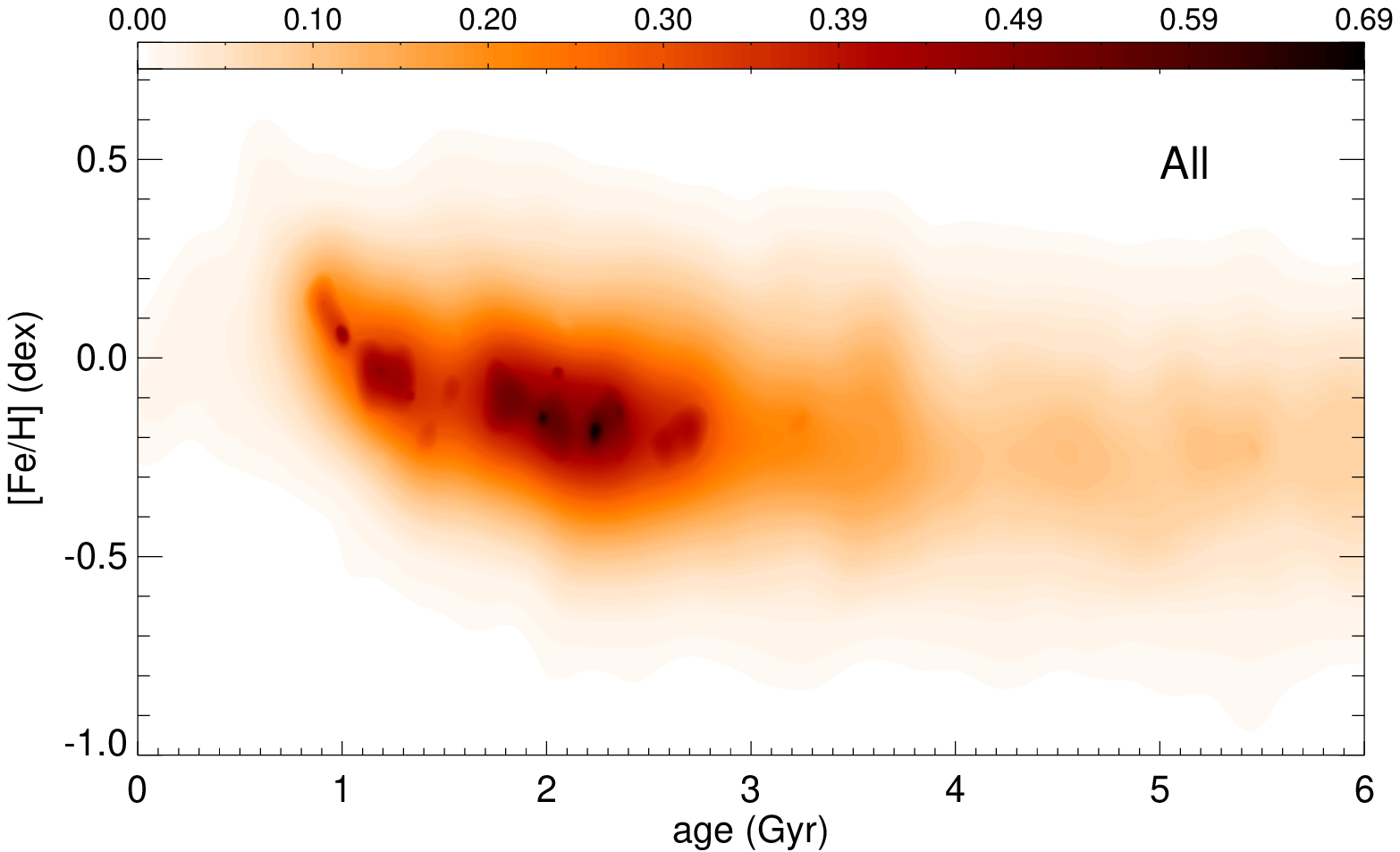}
       \includegraphics[width=0.35\textwidth]{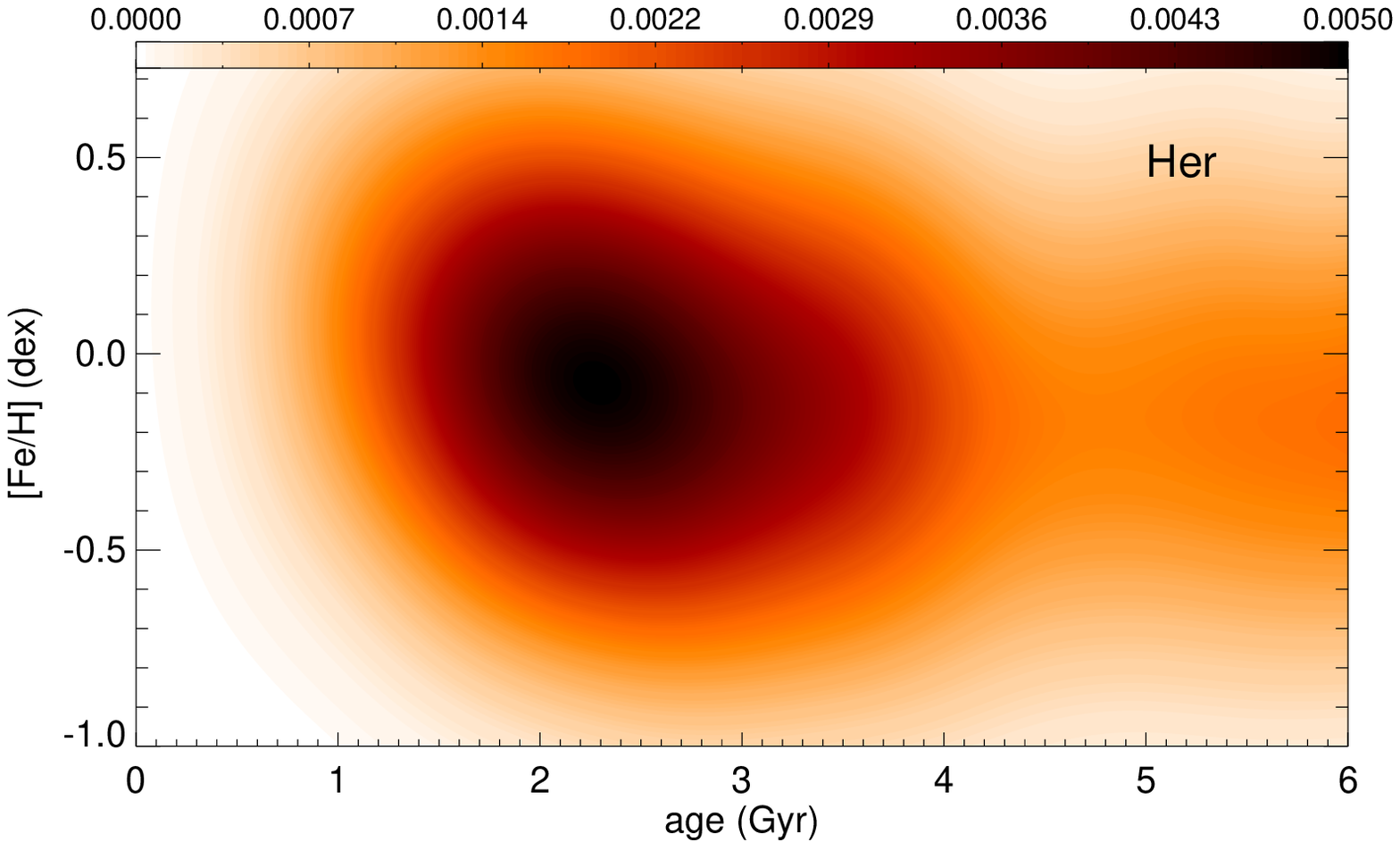}
       \includegraphics[width=0.35\textwidth]{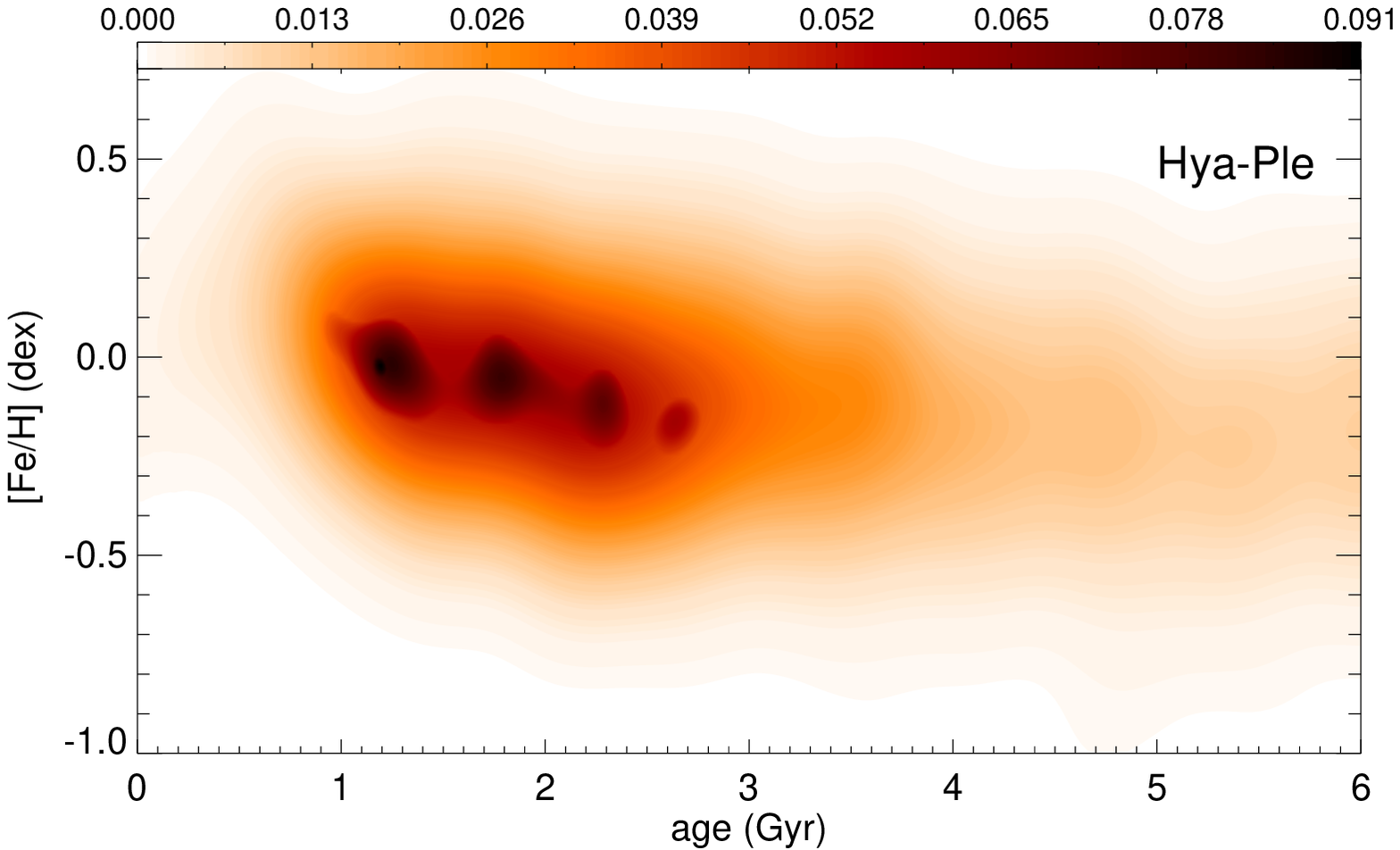}
       \includegraphics[width=0.35\textwidth]{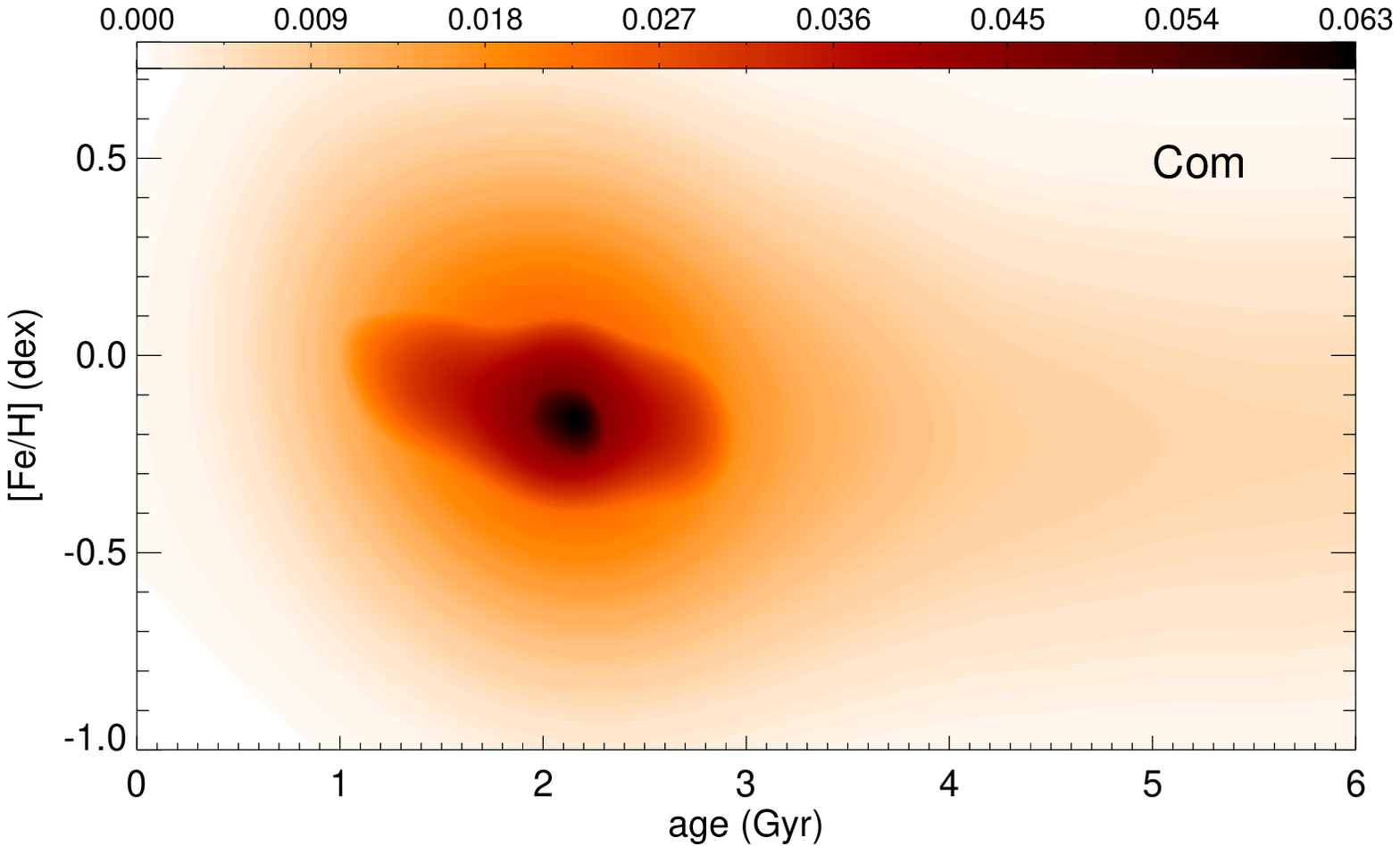}
       \includegraphics[width=0.35\textwidth]{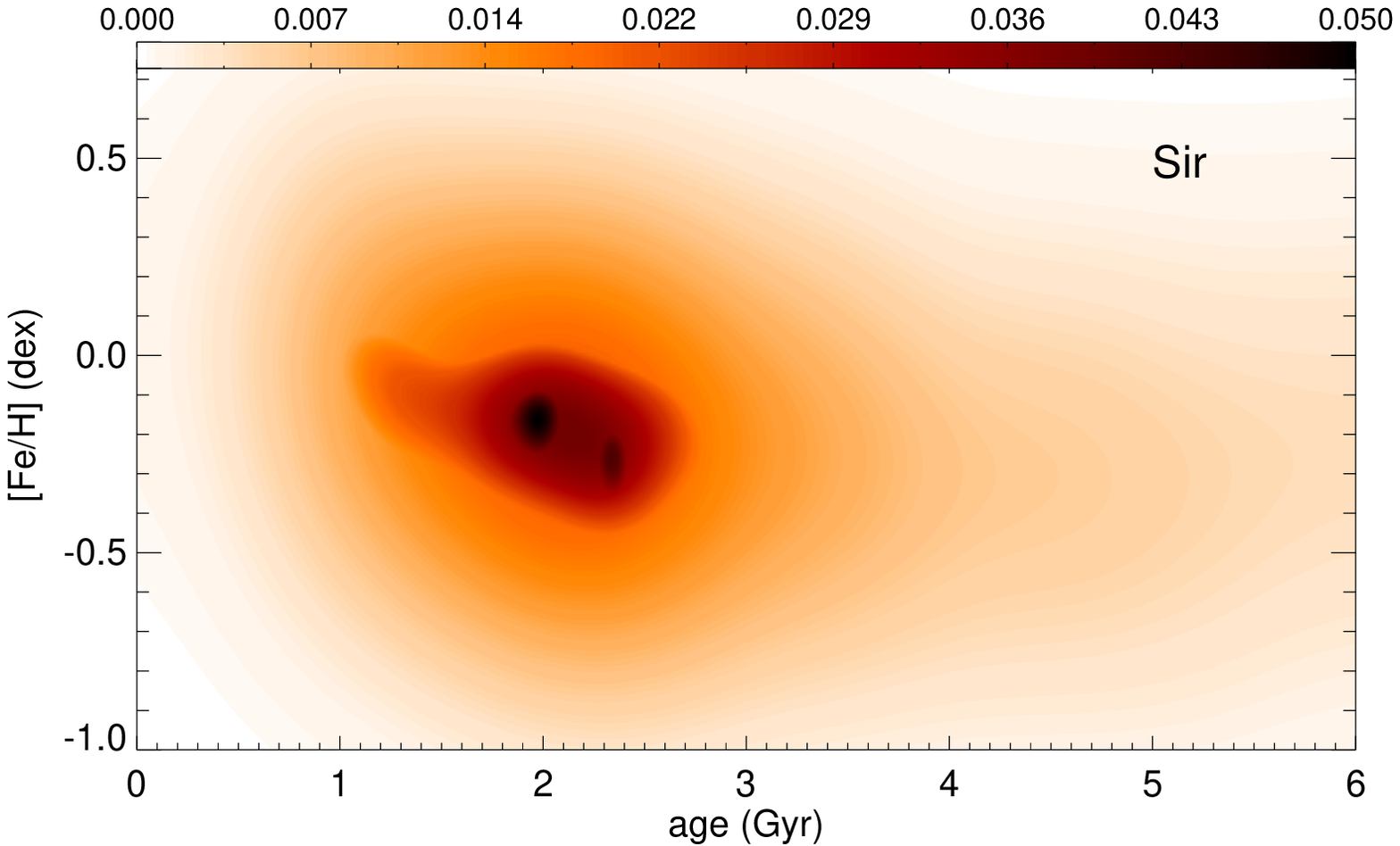}
 	\caption{Density field in the $age$--$[Fe/H]$ plane for all stars with available metallicities and ages (11215 stars) and for the different branches of Hercules, Hyades-Pleiades, Coma Berenices and Sirius (436, 1973, 1559, 1229 stars respectively) obtained by WD.}
      \label{agefeh}
\end{figure}

Figure \ref{agefeh} (top) shows the distribution of the whole sample in the $age$--$[Fe/H]$ plane. Due to the complexity in the age-metallicity relation (see the exhaustive discussion on the biases in \citealt{haywood06}) any new conclusion is beyond the scope of this paper. Our analysis is restricted to a comparison between branches. Figure \ref{agefeh} shows the distributions in the $age$--$[Fe/H]$ plane for the stars along each of the four branches. To a first approximation these distributions exhibit the same general tendency as the whole sample. The WD treatment allows us to detect other significant features such as the clumps in age for the Hyades-Pleiades branch detected in Sect. \ref{age} which show a decrease in $[Fe/H]$ with increasing age. For the range 1-3$\Gyr$, where more data is available, although it is preliminary, the slope of the Sirius distribution seems to be slightly more negative than for the other two main branches (from $\sim -0.10\pm0.01 dex/\Gyr$ for Sirius to $\sim -0.07\pm0.01 dex/\Gyr$) for Hyades-Pleiades.

\section{Summary and discussion} \label{conclusions}

Moving groups are becoming a powerful tool for studying the large-scale structure and dynamics of the Milky Way. We apply multi-scale techniques --wavelet denoising-- to an extensive compendium of more than 24000 stars in the solar neighbourhood to characterize the observed kinematic structures in  $U$--$V$--$age$--$[Fe/H]$ 4-dimensional space. In this paper we focus our analysis on establishing the observational constraints that will allow us to study the origin and evolution of these structures.

 The advent of Hipparcos astrometric data led to the definitive recognition of a non-smooth distribution function of the Galactic disc. Our results corroborate this and go one step further towards characterizing the velocity distribution function. Branches connecting the classic moving groups in the $U$--$V$ plane are definitely the dominant structures. We confirm the existence of the Sirius, Coma Berenices, Hyades-Pleiades and Hercules branches. The first three branches are spaced at intervals of approximately 15 $\kms$ with no significant variations with age or spectral type. The Hercules branch is located about 30 $\kms$ from the Hyades-Pleiades branch. The four branches present a negative slope of $\beta \sim 16\deg$ in the $U$--$V$ plane, lower than that found by \citet{skuljan99}. We have studied the density drops in the $U$--$V$ plane and the existence of abrupt edge lines is corroborated, especially at low $U$ and high $V$ and between the Hyades-Pleiades and the Hercules branches. Each branch presents a different density distribution in its extremes near these edge lines. These features definitely rule out the classic idea of a smooth velocity field distribution. The use of the same statistical method for the samples with different spectral types has made possible an exhaustive comparison between their kinematic planes. The branches are present in all the distributions. The study of variations of the kinematic structures with Galactic position has been presented for the KM giant sample. A significant change of contrast among substructures inside the branches is confirmed and the shape of the Hercules branch changes among regions, being more conspicuous in the region towards the anti-rotation direction with respect to the centre of this sample.

We have analysed, for the first time, the age and metallicity distributions of the branches. We confirm an extended age distribution for all the branches  but a different minimum age of their stars. The set of the three main branches of Hyades-Pleiades, Coma Berenices and Sirius is well established for stars $>400 \Myr$ and the extended branch-like shape of Hercules is detected in all subsamples with ages $> 2 \Gyr$.  The relative density between the Hyades-Pleiades and the Hercules branches clearly decreases with age. We find a periodicity in age of about 500-600 $\Myr$ in the Hyades-Pleiades branch. For the other branches only an outline of the shape of the whole age distribution is observed. A wide range of metallicity is found for each branch, especially for Hercules with a higher metallicity dispersion. A complex relation between kinematics and metallicity has been established.  Concerning the age-metallicity relation, differences are observed between branches. The periodic bumps in the age distribution for the Hyades-Pleiades branch show a decrease in metallicity with increasing age.

All the above observational results and, moreover, the whole set of distributions in the $U$--$V$--$age$--$[Fe/H]$ space presented here, are the fundamental elements necessary to check the present and future dynamic models proposed for the formation of kinematic structures. While some of these current models already explain some of the observational results, other important observational features remain unexplained. For instance, the simulations of \citet{desimone04} produce branches with a slope which fits that measured here. Furthermore, the model in \citet{quillen05} agrees with the fact that the Hyades-Pleiades branch is more metallic than the Coma Berenices branch. However, features such as the mix of stars in the Hyades and Pleiades moving groups in periodic clumps in age in the same branch --perhaps reflecting  a common mechanism for the formation of these two kinematic structures-- or the peculiar metallicity distribution inside branches demand an in-depth explanation. 

To make further progress in this field, simulations obtained from orbit integration under a model for the Galactic potential are being undertaken (Antoja et al. 2008, in preparation). After checking that the simulated kinematic structures critically depend on the gradients of the spiral arm potential, two different models for the Galactic spiral arms are being explored: the classic cosine perturbation \citep{lin71} and a 3D mass distribution (superposition of the inhomogeneous oblate spheroids in \citealt{pichardo03}). We are also incorporating stellar ages and metallicities in the simulations to trace the origin and evolutionary state of the test particles, which is another key element. Of course, other ingredients such as the Galactic bar will also be included to analyse the kinematic structures in an overall dynamic context. 

Thus, the present detection and characterization of the branches in $U$--$V$--$age$--$[Fe/H]$ 4-dimensional space provide us with the main properties summarized above. It is the whole set of distributions in this space that contains all the information. The application of the same statistical technique to both the observed and  simulated data will allow a direct comparison between them, and thus be a powerful test of the models for the formation of the structures. This comparison will eventually offer a physical interpretation of the formation of the kinematic structures in terms of local or large-scale dynamics.
\begin{acknowledgements}
This work was supported by the CICYT under contracts AYA2003-07736 and AYA2006-15623-C02-02. TA was supported by the Predoctoral Fellowships of the Generalitat de Catalunya 2006FI 00798 and 2007FIC 00687. We wish to thank J.L. Starck, F. Murtagh and V. Mart\'inez for providing us with the MR software packages and for their helpful comments. We also wish to thank J. Palou\v s for interesting discussions in Barcelona and C. Fabricius for the final reading of the manuscript. We would like to thank the referee for the very helpful comments and suggestions which much improved the first version of this paper.
\end{acknowledgements}

\end{document}